\documentclass[useAMS,usenatbib]{mnras}

\usepackage{mathtext,amssymb,amsmath}
\usepackage{epsfig}
\usepackage{graphics}
\usepackage{url}
\usepackage{adjustbox}

\usepackage{times}

\usepackage[T1]{fontenc}

\newcommand{\be}{\begin{equation}}
\newcommand{\ee}{\end{equation}}
\newcommand{\beq}{\begin{eqnarray}}
\newcommand{\eeq}{\end{eqnarray}}

\def\ergcm2s{{\rm erg\,cm^{-2}\,s^{-1}}}

\title[Intermediate polars in Gaia era]{Hard X-ray view on intermediate polars in the Gaia era}
\author[V. F. Suleimanov et al.]{Valery F. Suleimanov,$^{1,2,3}$\thanks{E-mail: suleimanov@astro.uni-tuebingen.de}
Victor Doroshenko,$^{1}$
and Klaus Werner$^{1}$
\\
$^{1}$Institut f\"{u}r Astronomie und Astrophysik, Kepler Center for Astro and
Particle Physics, Universit\"{a}t T\"{u}bingen, Sand 1, 72076 T\"{u}bingen,
Germany\\
$^2$Kazan (Volga region) Federal University,  Kremlevskaya str. 18, Kazan 420008, Russia\\
$^3$Space Research Institute, Russian Academy of Sciences, Profsoyuznaya 84/32,
117997 Moscow, Russia}

\pubyear{2018}

\begin{document}
\label{firstpage}
\pagerange{\pageref{firstpage}--\pageref{lastpage}}
\maketitle

\begin{abstract}
The hardness of the X-ray spectra of intermediate polars (IPs) is
determined mainly by the white dwarf (WD) compactness (mass-radius
ratio, $M/R$) and, thus, hard X-ray spectra can be used to constrain
the WD mass. An accurate mass estimate requires the finite size of the
WD magnetosphere $R_{\rm m}$ to be taken into the account. We
suggested to derive it either directly from the observed break
frequency in power spectrum of X-ray or optical lightcurves of
a polar, or assuming the corotation. Here we
apply this method to all IPs observed by NuSTAR (10 objects) and
Swift/BAT (35 objects). For the dwarf nova GK\,Per we also observe a
change of the break frequency with flux, which allows to constrain the
dependence of the magnetosphere radius on the
mass-accretion rate. For our analysis we calculated an additional grid of
two-parameter ($M$ and $R_{\rm m}/R$) model spectra assuming a fixed,
tall height of the accretion column $H_{\rm sh}/R=0.25$, which is
appropriate to determine WD masses in low mass-accretion IPs like
EX\,Hya. Using the Gaia Data Release 2 we obtain for the first time
reliable estimates of the mass-accretion rate and the magnetic field
strength at the WD surface for a large fraction of objects in our
sample. We find that most IPs accrete at rate of
$\sim10^{-9}$\,M$_\odot$\,yr$^{-1}$, and have magnetic fields in the
range 1--10\,MG. The resulting WD mass average of our sample is
$0.79\pm0.16$\,M$_\odot$, which is consistent with earlier estimates.
\end{abstract}



\section{Introduction}

Intermediate polars (IPs) are a subclass of cataclysmic variables
(CVs), accreting close binary systems with white dwarfs (WDs) as
accretors and main sequence dwarfs, overfilling their Roche lobe, as
donors \citep[see details in][]{Warn:03}. The WDs in IPs are
magnetized enough (0.1--10\,MG) to destroy the accretion disc at some
distance (the magnetospheric radius $R_{\rm m}$), but not enough to
push the magnetosphere beyond the corotation radius $R_{\rm c}$ and
inhibit the accretion. Here $R_{\rm c}$ is defined as the radius where the
Kepler frequency in the accretion disc equals the spin frequency of
the WD. At smaller radii, accreting material falls along magnetic
field lines onto the WD surface forming a standing shock above the
surface, where the kinetic energy of the falling matter is transformed
into thermal energy, and radiated away as optically thin plasma emission with
typical temperatures of a few tens keV \citep{Aizu:73, FPR:76}. The
plasma thus cools by radiation and settles in a subsonic regime
below the shock. This region is usually referred to as a post-shock
region (PSR) or an accretion column.
 
Hard X-ray radiation of a PSR allows to evaluate the WD mass $M$,
because the post-shock temperature is determined mainly by the
compactness of the WD \citep{Katz:77, Rothschild.etal:81}. This idea
was further developed and exploited to estimate masses of several WDs
\citep{Wu.etal:94, Cropper.etal:98, Cropper.etal:99, Ramsay:00}. These
authors developed the PSR theory to very high sophistication including
cyclotron cooling, the dipole geometry of PSRs, and the possible
difference between electron and proton temperatures
\citep{Canalle.etal:05, Saxton.etal:07}. In these works, however,
X-ray spectra in the classical X-ray band ($< 20$\,keV) were
considered.  This resulted in large uncertainties for the determined
masses as the Wien exponential cutoff of the PSR spectra, which is
essential for an accurate determination of the temperature, occurs at
higher energies.

More reliable results on WD masses in IPs were obtained after the
launch of X-ray observatories designed for the hard X-ray band. Mass
determinations using the PSR model spectra were performed using data from
RXTE/HEXTE \citep{SRR:05}, Swift/BAT \citep{Betal:09}, and Suzaku
\citep{Yuasa.etal:10, HI:14b, Yuasa.etal:16} observatories. Several IPs were
investigated also using INTEGRAL/IBIS observations
\citep{Revnivtsev.etal:04b, Falanga.etal:05}, however, most of them
were interpreted assuming generic hot optically thin plasma radiation
models rather than PSR models calculated from first principles
\citep{Barlow.etal:06, Landi.etal:09, Bernardini.etal:12}. Given the
importance of the cutoff for WD mass measurements, and comparatively
faintness of IPs in the hard band, currently the data provided by the
NuSTAR observatory \citep{NuSTAR} suit best for this purpose. Several
IPs were studied by NuSTAR using PSR model spectra
\citep{Hailey.etal:16, Suleimanov.etal:16,Shaw.etal:18,
  Wada.etal:18}. A recent review of hard X-ray emitting IPs was
presented by \citet{Mukai:17}.

The commonly used method for mass determination of WDs in IPs described above
is not without flaws. There are several theoretical uncertainties which can
affect the results, most notably the assumed geometry of the PSR and the finite
but not well defined magnetospheric radius. For the PSR geometry the dipole
approximation appears to be well justified, and an accurate theory to calculate
its structure has been developed by \citet{Canalle.etal:05}. A simplified
approach has been suggested by \citet{HI:14a}, where the dipole geometry is
approximated by a PSR cross-section increasing with distance from the WD
surface.

Using of this simplified approach allowed \citet{Suleimanov.etal:16} to take
also the finite magnetosphere size into account. The latter development is
particularly important for objects with relatively small magnetospheres with
radii of a few WD radii $R$. Matter starting to fall from a small radius
accelerates to significantly smaller velocities compared to the case of
accretion from infinity. As a result, the WD mass might be significantly
underestimated if this effect is not accounted for \citep{SRR:05}. Whether the
assumed magnetospheric radius affects the resulting WD mass estimate is not
known a priori, because that PSR modeling of the X-ray spectra only allows to
constrain $M-R_{\rm m}/R$.

\citet{Suleimanov.etal:16} computed a two-parameter ($M$ and $R_{\rm
  m}/R$) grid of PSR model spectra in order to obtain such constraints
from observations. To break the degeneracy and estimate the mass of
the WD, an independent estimate of the magnetosphere size is
required. \citet{Suleimanov.etal:16} suggested to use the break
frequency in the power spectrum of the studied IP assuming that it
corresponds to the Keplerian frequency of the disc at the
magnetospheric radius \citep[][]{Revnivtsev.etal:09,
  Revnivtsev.etal:11}. The method was applied to two IPs with small
magnetospheres, EX\,Hya and GK\,Per, and indeed it was found that
under this assumption the resulting mass estimates better agree with other estimates. For
GK\,Per, which was observed in different luminosity states, we were
also able to investigate the dependence of the magnetospheric radius
on the mass-accretion rate \citep{Suleimanov.etal:16}. 

Another theoretical uncertainty is related to the generally unknown
local mass-accretion rate $a$\,(g\,s$^{-1}$\,cm$^{-2}$). Indeed, PSRs
with low $a$ and, therefore, low plasma density are expected to cool
slowly and thus a larger shock height can be expected. This reduces
the free-fall velocity at the shock, i.e., it might also lead to an
underestimation of the WD mass. \citet{Luna.etal:18} mentioned the
significance of the PSR height for the low-luminosity IP EX\,Hya. In this
paper we present a new two-parameter grid of PSR model spectra where
this effect is quantified, and apply it to several IPs accreting at
low rate.

The spectra of some luminous IPs, such as V1223\,Sgr, have a very
complicated structure at low photon energies ($< 10$\,keV), presumably
due to complex absorption within the binary system or/and
reflection off the WD's surface \citep[see, e.g.][]{Cropper.etal:98,
  Cropper.etal:99, SRR:05}.  Commonly, a partially covering absorber
is considered to model the  observed spectra \citep[see,
  e.g.,][]{Cropper.etal:99}. Indeed, IPs rotate and the radiating PSR
is expected to be periodically occulted by material falling along
magnetic force lines (accretion curtains).  As a typical observation
duration is much longer than the IP spin period, part of the time only
strongly absorbed emission from the PSR is observed. This effect is
expected to be more pronounced for systems accreting at high rate
where the density of the accretion flow is higher, which does indeed
seem to be the case.

An additional aspect to explain the observed spectra is to include the
reflection component \citep{Cropper.etal:98}. This scenario is also
physically motivated because the PSR radiation has to be reflected
from the WD surface. Moreover, the iron-line complex at 6--7\,keV
observed in the majority of IPs is more readily explained in this
scenario. We note, however, that the transmission of hard X-ray
radiation through cold matter will also generate an iron fluorescence
line. Moreover, the contribution of $K_\alpha$ is small in the spectra
of low-luminosity IPs \citep[e.g., EX\,Hya][]{Luna.etal:15} and large
(more than a half of the total equivalent width of the complex) in the
high-luminosity IPs like GK\,Per during outburst
\citep{Yuasa.etal:16}.  

On the other hand, the importance of reflection was recently established for some IPs
using NuSTAR observations \citep{Mukai.etal:15}, and proved to be an
important factor to be taken into account when determining the WD mass
using broadband spectra \citep{Shaw.etal:18}. We note, however, that
detailed calculations of the reflected spectrum taking the geometry of
the PSR into the account are required for such estimates to be robust,
and no such calculations are available as of now. Furthermore, the
relative importance of partially covering absorption and reflection
components is still unknown and ignoring either effect could bias the
mass estimates. In the present work we restrict, therefore, the
analysis to hard X-rays ($> 20$\,keV) where both effects are
less prominent (\citealt{Suleimanov.etal:16}, see also discussion in
\citealt{Hailey.etal:16}).

Generally, it is important to emphasize that the knowledge of accurate
WD masses in IPs is not only essential to understand the physics of
individual objects, but also to understand the composition and
evolution of the Milky Way as a whole. Indeed, the IP population is
substantial, and it is known to be responsible for a large fraction of
the total hard X-ray emission of the Galaxy and other galaxies
\citep{Muno.etal:04,Krivonos.etal:07, RNat:09, Yuasa.etal:12}. The
spectra of IPs depend on their WD mass, so the average WD mass in the
IP population is an important parameter to constrain the properties of
an unresolved IP population.

First estimates of the average WD mass in the IP population in the
Galaxy ridge and bulge based on modeling of the observed hard X-ray
luminosity function gave relatively low masses of about
0.5--0.7\,M$_\odot$ \citep[see, e.g.][]{Yuasa.etal:12}. On the other
hand, a recent investigation using NuSTAR observations found an
average WD mass close to the value typical for the nearby CV
population \citep[$\sim$\,0.9\,M$_\odot$,][]{Hailey.etal:16}.

The average mass of WDs in CVs is known to be larger
\citep[$\sim$\,0.8\,M$_\odot$, ][]{Zorotovic.etal:11} than that of field
WDs \citep[$\sim$\,0.6\,M$_\odot$, ][]{Kepler.etal:16}. The same is true
for nearby IPs, where the average mass was estimated to be in the
range of 0.8--0.9\,M$_\odot$ \citep[e.g.,][]{Yuasa.etal:10,
  Bernardini.etal:12}. Several possibilities to explain the difference
were discussed \citep[see, e.g.,][]{Zorotovic.etal:11}, however, any
meaningful conclusions require that the comparison is done for as
large a uniform sample of WD masses in IPs as possible. Obviously it
is also important that all known systematic effects which could bias
the mass estimates are considered. In the current work we attempt to
provide such a sample by analyzing all available observations of IPs
performed by the NuSTAR and Swift/BAT observatories. Distances to
close IPs are known with high accuracy from the Gaia Data Release 2
\citep{Gaia_mission,Gaiadr2}. This offers the possibility to also
obtain reliable estimates for mass-accretion rates and magnetic field
strengths for a large fraction of the objects in our sample.

\section{Updated model of the PSR spectrum}

As discussed above, the assumed size of the IP magnetosphere directly
influences the observed PSR spectrum and thus the WD mass estimation. To
account for this quantitatively, we have recently developed a
two-parameter grid of hard X-ray model spectra of IPs
\citep{Suleimanov.etal:16} which includes the magnetospheric radius
expressed in units of WD radius as the second parameter, besides the WD
mass. The hardness of the PSR spectrum (or the maximum temperature of the PSR)
depends on the velocity ${\rm v}_{\rm ff}$ of the accreting matter \citep[see, e.g.][]{FKR02}
\be
       kT_{\rm sh} = \frac{3}{16}\mu m_{\rm H}\,{\rm v}^2_{\rm ff},
       \ee
where $\mu\approx 0.62$ is the mean molecular weight of a fully ionized
plasma with solar chemical composition, $k$ is Boltzmann constant,
and $m_{\rm H}$ is the proton mass. The velocity depends not
only on the WD's $M$ and $R$, but also on the initial
potential energy of the accretion flow, and height of the PSR, $H_{\rm
  sh}$. The potential energy is defined by the height from which matter
starts to fall freely, which to first order must be comparable
to the magnetosphere size $R_{\rm m}$
\be
      {\rm v}^2_{\rm ff} = 2GM \left(\frac{1}{R+H_{\rm sh}} - \frac{1}{R_{\rm m}}\right).
\ee     
The height of the shock $H_{\rm sh}$ depends mainly on the local mass-accretion rate $a$,
and becomes important only at low
rates, i.e., $a < 1$\,g\,s$^{-1}$\,cm$^{-2}$.

 \begin{figure} 
\begin{center}  
\includegraphics[width= 1.\columnwidth]{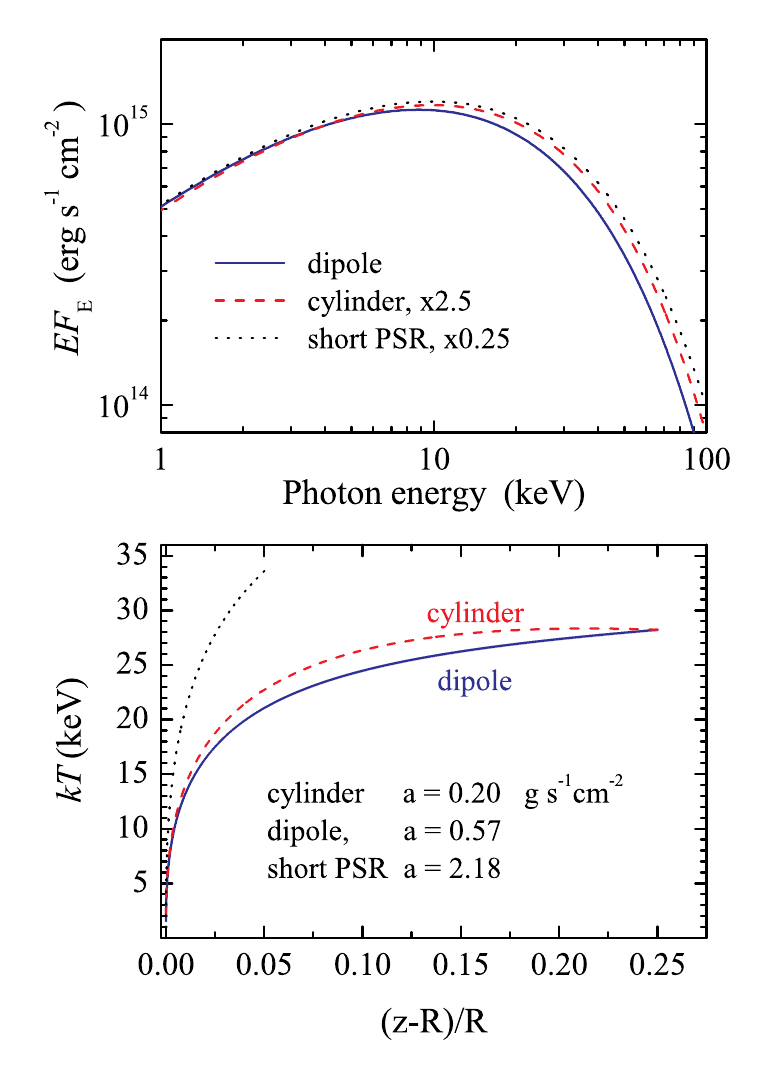}
\caption{\label{fig:dip_tall} 
Comparison of model spectra (top panel) and temperature structures (bottom panel)
of short and tall PSRs. The tall PSR model is computed for two geometries (cylinder and dipole).
} 
\end{center} 
\end{figure}

In our previous work \citep{Suleimanov.etal:16} we only considered the
case of high accretion rates, which is relevant for most IPs. However,
the respective model grid could be inadequate for IPs with low
luminosities where the local mass-accretion rate might be low and thus
the height of the shock must be taken into account. For instance, the
closest known IP, EX Hya, has a very low luminosity, $L \approx
10^{32}$\,erg\,s$^{-1}$\,cm$^{-2}$, see, e.g.,
\citet{SRR:05}. Recently \citet{Luna.etal:18} concluded that the PSR
in this IP could indeed be tall enough to affect the spectrum. As we
will discuss below, the situation is similar for several other
sources, so we extend the model presented in
\citet{Suleimanov.etal:16} also to low accretion rates.

Given the uncertainties in our knowledge of the local accretion rates,
we only computed a single two-parameter model grid for fixed $H_{\rm
  sh}/R=0.25$, which corresponds to the tall-column limit. We
considered WD masses $m=M/$M$_\odot$ from 0.3 to 1.2 equidistantly
distributed within this range with a step of 0.02. The magnetospheric
radius is expressed in units of the WD radius, $r_{\rm m} = R_{\rm
  m}/R$ and changes as $60/N$, where $N$ takes values from 1 to 40. In
addition we computed models for very large $r_{\rm m} = 1000$  to
simulate accretion from infinity. Details of our model computations
are described in \citet{Suleimanov.etal:16}. The main feature of the
method is using a quasi-dipole geometry of the PSR as suggested by
\citet{HI:14a}.  Effectively we assume that the PSR cross-section
increases with height above the WD surface (defined along the vertical
coordinate $z$) as $S=(1+z/R)^3$ \citep{HI:14a}. The only difference
between our models presented earlier and here is, thus, the assumed
local mass-accretion rate. In \citet{Suleimanov.etal:16} we fixed the
total mass-accretion rate to $\dot M = 10^{16}$\,g\,s$^{-1}$ and the
fraction of the WD surface occupied by the PSR to $f=S_{\rm PSR}/4\pi
R^2=5\times10^{-4}$. The local accretion rate $a$ was then adjusted to
match these values for given WD mass and radius. In contrast, here we
consider the tall-column limit and fix the PSR height to $H_{\rm
  sh}/R=0.25$, again adjusting the local accretion rate $a$ to match
this value for given WD mass and radius. 

The impact of increased accretion-column height on temperature
structure and emerging spectrum is illustrated in
Fig.\,\ref{fig:dip_tall}. As expected, the spectrum of a tall PSR is
softer than that of a short PSR. This effect is more significant
for a PSR model assuming quasi-dipole geometry in comparison with a
PSR model with simple cylindrical geometry. The reason is that the
density of the model computed in quasi-dipole geometry is higher,
which increases the cooling rate compared to the cylindrical model.

Speaking of cooling, it is important to emphasize that while we use an
accurately computed cooling function to calculate the PSR structure,
only free-free emission assuming the full ionization of 15 chemical
elements is used to calculate the emerging spectra \citep[see details
  in][]{Suleimanov.etal:16}. Therefore, the models underestimate the
bolometric fluxes compared to actual PSR luminosities. Quantitatively,
this reduction slightly depends on the assumed magnetosphere size and
mass of the WD, and as a result the model spectra give bolometric
luminosities of about 70--80\% of the total PSR luminosities. This
does not affect shape of the spectrum or derived WD mass, because
almost all the missing radiation escapes below 3~keV, but it is
important for estimates of the accretion rate based on the observed
flux. 

\section{Observations and data analysis}
\label{s:ctm}
To confront models with observations, we opted to use two
data sets. The first consists of dedicated IP observations carried out
by the NuSTAR observatory either individually, or in the framework of the
currently ongoing legacy survey for IPs. In the range 20--80~keV,
NuSTAR has the highest sensitivity of all past and currently operating
facilities, and thus is particularly well suited for studies of faint
sources, such as WD
mass measurements. The IPs observed by NuSTAR together with effective
exposure times are listed in Table~\ref{tab:nuobs} (see also Table\,\ref{tab1}). To complement the
NuSTAR dataset we used IP spectra provided by the Swift/BAT
105-Month Hard X-ray Survey \citep{2018ApJS..235....4O}. Despite the lower
sensitivity, long cumulative exposure across the entire sky allowed
to significantly increase the sample of considered sources. It is
worth noting that spectra in this dataset are integrated over a 105
months period, so the results are not really meaningful for transient
sources like GK\,Per, where dedicated observations are required. The
IPs from the Swift/BAT Survey considered in this work are
listed in Table~\ref{tab2}.

\begin{table}[!b]
\centering   
\caption{List of IPs observed by NuSTAR and considered in this work. Effective exposures times after standard screening are also listed. 
 \label{tab:nuobs} 
 }
\begin{tabular}[c]{l c c }
\hline
Name & 	Obs. id       &  Exposure, ks         \\ 
\hline 
NY Lup		        &   30001146002 	 	&  23\\
EX Hya			&    30201016002 		&  25\\
V2731 Oph 		&  30001019002 		&  49 \\ 
GK Per	               & 90001008002 		&  42\\
		& 30101021002 		&  72\\
TV Col	& 30001020002 & 50\\
V1223 Sgr 		& 30001144002		& 20\\
V405 Aur 			&  30460007002 		&38 \\
RX\,J2133+5107 	& 30460001002	 	& 26 \\
FO Aqr 			&  30460002002 		&  26\\
V709 Cas			&  30001145002 		&  26\\
 \hline
\end{tabular}
\end{table}

As already mentioned in the introduction, many IPs have rather complex
spectra below 10\,keV either due to the presence of a partial covering
absorber or due to reflection. While we have found that it is possible
to adequately describe broadband NuSTAR spectra in the 3--80\,keV range
with PSR models in combination with either a partial covering absorber
or a reflection component, or both \citep[see also][]{Shaw.etal:18},
we found also that fit results including the deduced WD mass are
dominated in this case by the soft band, where most photons
are detected, and thus strongly affected by the quantitative
description of the absorption/reflection modifying the PSR
spectrum. On the other hand, the WD mass in the model is
effectively defined by the high-energy rollover which is generally
above 20\,keV. Given that no robust description of the
absorption/reflection exists, we chose, therefore, to restrict the
analysis to $E>20$\,keV for NuSTAR and $E>15$\,keV for Swift/BAT,
where the contribution of these effects are less important. The
advantage of this approach is that a much simpler model can be used to
describe the spectra in this case. In particular, we only use a PSR
model component, which in the end implies that final statistical
uncertainties for the derived WD parameters are comparable to the case
when the entire energy band is described with a more sophisticated
model. Another advantage is that we can also directly compare results
obtained with both instruments as they effectively operate in same
energy range.

%

For the reduction of the Swift/BAT data, we used survey-averaged
response files, and assigned conservative 10\% model systematics for
all sources to account  for deviations from the model observed in
first energy bin of some objects, presumably due to
absorption/reflection which might be still significant in 15--20\,keV
band  \citep[see also][]{Shaw.etal:18}. We opted for this approach to
ensure that an acceptable fit can be achieved for all objects in a
uniform way.

The reduction of the NuSTAR data was carried out using the
\textit{heasoft 6.24} package and current calibration files
(v20180814). The source spectra were extracted for each of the two
NuSTAR modules independently from source-centered regions with radius
of 45--80$''$. The extraction radius was optimized individually for
each source in order to achieve best signal-to-noise ratio in the
20--80\,keV band. Background spectra were extracted from the adjacent
source-free regions for each observation. All spectra were grouped to
contain at least 25 source counts per energy bin. For each observation
the spectra were extracted and modeled for both NuSTAR
modules independently, with all fit parameters except normalization
linked for both modules. No systematic error was included in this case.
Note that for plotting we combine the data from both NuSTAR units, and group
the combined spectra to contain at least 100 counts per energy bin to enchance
the clarity of the figures.

We have also carried out a timing analysis for all IPs observed by
NuSTAR with the aim to detect the break in the aperiodic power
spectrum and to constrain the break frequency. To increase
counting statistics, light curves with time bin of 10\,s in the full
3--80\,keV band were extracted, with light curves from the two modules
co-added. The light curves were then corrected to the solar
barycenter. No background subtraction was done because background is
negligible in this case, and furthermore, not relevant for the timing
analysis. Besides the aperiodic variability, IPs exhibit also coherent
pulsations which need to be subtracted prior to modeling of the
aperiodic power spectrum. To subtract the pulsations, we followed the
procedure suggested by \citet{Revnivtsev.etal:09}. For each
observation we first obtained average folded pulse profiles using the
spin periods from literature (see Table~\ref{tab2}) and the
\textit{heasoft} task \textit{efold}. Using the same folding
parameters, we then calculated the pulse phase for each light curve
time bin, and subtracted the expected average rate for the respective
time bin, which effectively suppresses pulsations. The power spectra
were then obtained using the task \textit{powspec} and binned
logarithmically. 

For the sources where a break in the aperiodic power spectrum was
detected, we converted the obtained power spectrum into a format
readable by \textit{Xspec} using the \textit{heasoft} task
\textit{flx2xsp} \citep[see also][]{2012MNRAS.419.2369I}. Power
spectra were then modeled together with the energy spectra using a
broken power-law model with the break frequency linked to the 
magnetospheric radius parameter in PSR model, assuming that the break corresponds to
the Keplerian frequency at this radius.

 \begin{figure} 
\begin{center}  
\includegraphics[width=  0.9\columnwidth]{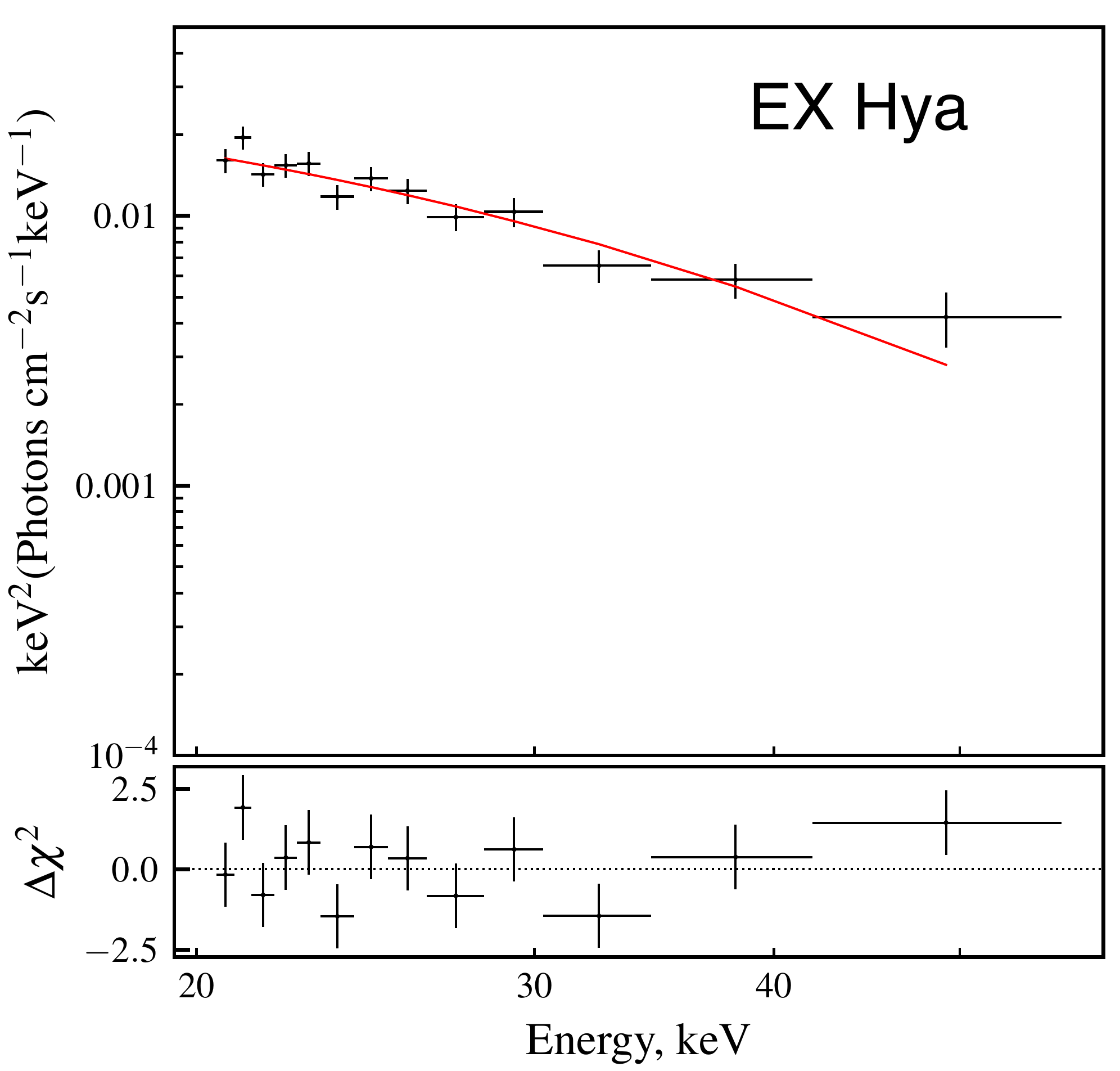}
\includegraphics[width=  0.9\columnwidth]{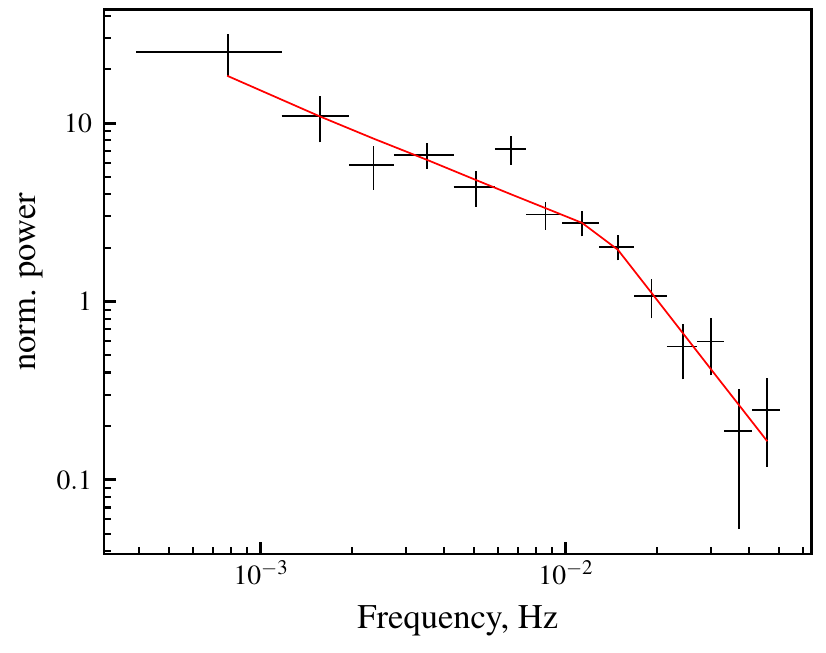}
\caption{\label{fig:exhya_sp}
Top panel: NuSTAR spectrum of EX Hya fitted by a tall PSR model ($M = 0.70$\,M$_\odot$ and $R_{\rm m} = 3.4\,R$) together with residuals of the fit. 
Spectra from both NuSTAR units combined for plotting to enhance clarity. Bottom panel:  the power
spectrum of EX\,Hya fitted with a broken power law, $\nu_{\rm br} = 0.013$.  
} 
\end{center} 
\end{figure}

 \begin{figure} 
\begin{center}  
\includegraphics[width=  0.9\columnwidth]{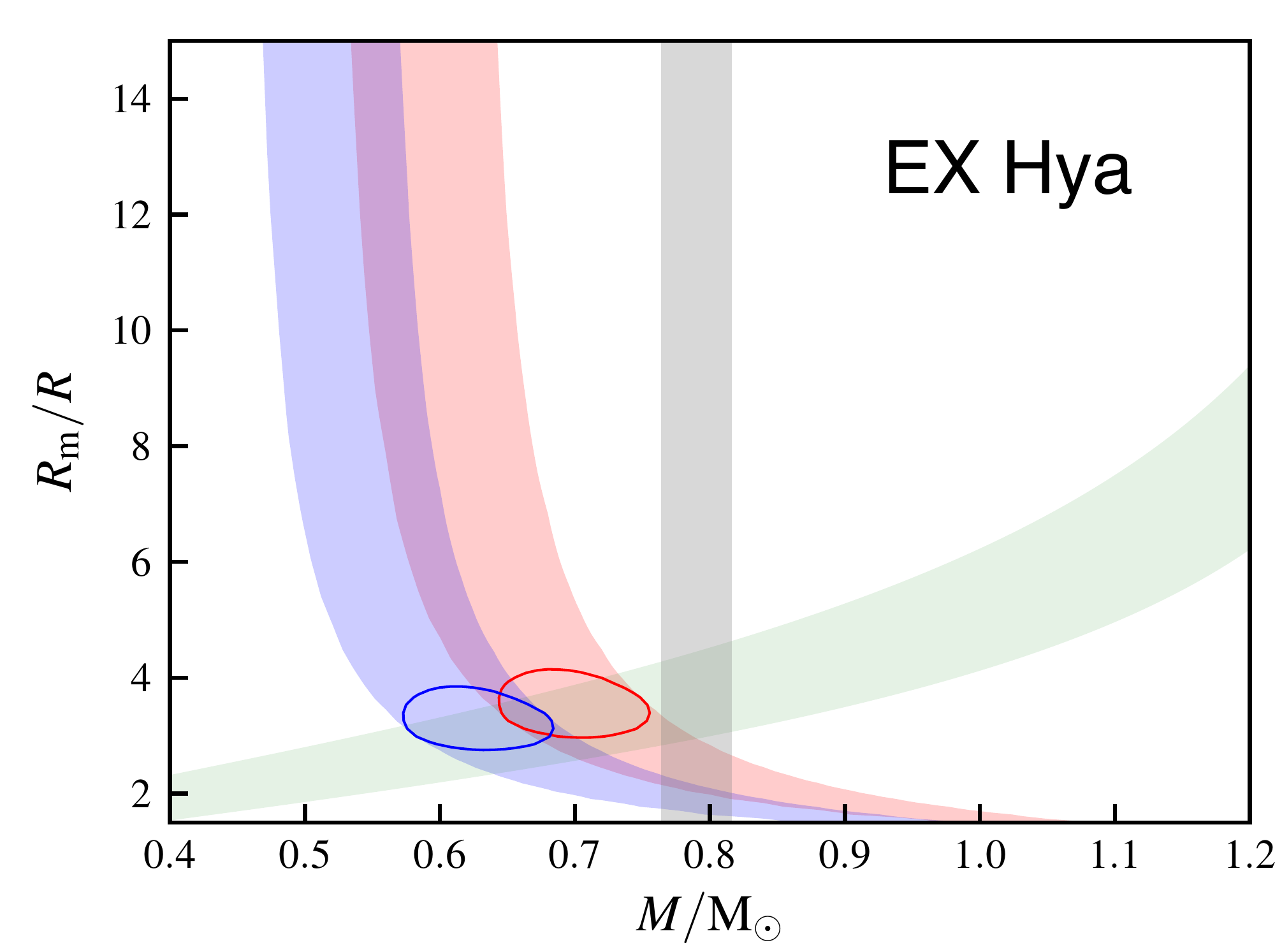}
\caption{\label{fig:exhya} 
Strips in the $m - r_{\rm m}$ plane obtained using spectral fitting by
short (blue strip) and tall (red strip) PSR models, and using 
the break frequency $\nu_{\rm br} = 0.013$\, Hz in the power spectrum of EX Hya (green strip). 
The  corresponding most probable regions are shown by the solid closed curves.
 The vertical strip corresponds to the WD mass derived from optical observations.} 
\end{center} 
\end{figure}

\subsection{NuSTAR observations of EX Hya}

As already mentioned, there is an intrinsic degeneracy of WD
mass and height from which the accretion flow starts to fall freely,
i.e. the magnetosphere radius. Therefore, the magnetosphere size must
be estimated independently. The simplest assumption one can make is to
assume that a given WD is close to corotation, i.e., the magnetospheric
radius is equal to the corotation radius 
\be
R_{\rm m}\simeq R_{\rm c}=\left(\frac{GMP_{\rm spin}^2}{4\pi^2}\right)^{1/3}
\ee
where $P_{\rm spin}$ is spin period of the WD. This
assumption is likely well justified for persistent sources, or
transients in quiescence, but must clearly be violated in some cases,
most notably for transients during outburst.

At higher accretion rates the magnetosphere is compressed by the
accretion flow. The size of the magnetosphere is defined mostly by the
magnetic field strength and accretion rate in this case.
\citet{Revnivtsev.etal:09} demonstrated that observed power spectra of
the stochastic variability in accreting systems with magnetized
compact objects exhibit a break at frequencies close to the compact
object's spin period. Furthermore, the frequency of the break was
found to be correlated with the accretion rate, and comparable to the
expected Keplerian frequency at the magnetosphere radius.

Later \citet{Revnivtsev.etal:10} and \citet{Revnivtsev.etal:11} used
this fact to estimate the magnetospheric radius in several IPs. In
particular, they found the break frequency for EX Hya, $\nu_{\rm br} =
0.021\pm 0.001$\,Hz \citep{Revnivtsev.etal:11}, which corresponds to
a magnetospheric radius of $R_{\rm m} \approx 2.7 R$ assuming that
the break frequency equals the Kepler frequency at the magnetospheric
radius and a WD mass of 0.79 M$_\odot$ deduced from optical
measurements \citep{BR:08}.

Generally the WD mass is unknown and the measured break frequency only
constrains a region in the $m - r_{\rm m}$ plane 
\be 
\nu_{\rm br}=\sqrt{\frac{GM}{2\pi R^3_{\rm m}}}.
\label{eq:br}
\ee 
The intersection of this
region with the region obtained through X-ray spectrum fitting can
then be used to estimate both mass and magnetospheric radius of a
given WD simultaneously. 
We applied this approach to obtain parameters
of two IPs, EX Hya and GK~Per \citep{Suleimanov.etal:16}. For EX Hya,
using Suzaku data, we found the break frequency $\nu_{\rm br}
=0.021\pm0.006$\,Hz, which together with spectral modeling yielded
$M=0.73\pm0.06$\,M$_\odot$ and $R_{\rm m} = 2.6\pm 0.4\,R$. These
values are in good agreement with estimates obtained by other
authors (see references above). Given that GK~Per was observed in two
luminosity states, we also investigated  the dependence of the
magnetosphere radius on luminosity.
 
Later on, EX Hya has been also observed by
NuSTAR \citep{Luna.etal:18}. Here we use this observation to verify
the results obtained previously as well as the hypothesis that the height of the
PSR might be sufficiently large to affect its spectrum
\citep{Luna.etal:18}. Compared to \citet{Suleimanov.etal:16}, we also
use the improved method to merge the two constraints. Instead of fitting
an observed spectrum and power spectra independently and then
determining the intersection between the constraints provided by both
measurements on the $m - r_{\rm m}$ plane, we now fit both the energy and power
spectra simultaneously.
%
%
The break frequency is not considered
as a free parameter but instead is linked to the $R_m$ parameter in the PSR
model, assuming that the break occurs at the Keplerian frequency for a
given radius, i.e. using Eq.~\ref{eq:br} and \ref{eq:rm}. This approach allows to properly take the statistical
uncertainties for both energy and power spectra into the account, and
to directly evaluate the intersection region in the $m - r_{\rm m}$ plane.
 
The results are presented in Figs.\,\ref{fig:exhya_sp} and
\ref{fig:exhya}. The best fit to the observed energy spectrum and
power spectrum is shown in Fig.\,\ref{fig:exhya_sp}. Only the fit
obtained with the tall PSR model is shown because the difference
between the fits obtained for both model grids is negligible and only
the derived mass is slightly different. This difference is illustrated
in Fig.\,\ref{fig:exhya}. Unlike most of the plots, we adopt a
$3\sigma$ confidence level to emphasize the significance of the
difference between the two model grids.  It is also clear that results of the fitting with the tall
PSR models better agree with the optical measurement of the WD
mass. Therefore, we confirm the suggestion by
\citet{Luna.etal:18} that a tall PSR in EX Hya might be the reason
behind the discrepancies between different mass estimates previously
reported. The WD mass and the magnetospheric radius deduced using the
tall PSR model are  $0.70\pm0.04$\,M$_\odot$ and $3.4^{+0.4}_{-0.3}
R$. Note that EX\,Hya might have an even taller PSR than assumed in the tall
column model, $H_{\rm sh}/R=0.25$.

 The distance to EX Hya is known
with high accuracy after Gaia DR2. After conversion of the
measured parallax we obtained $D =56.95 \pm 0.13$\,pc. Therefore, the
mass accretion rate in the system can be deduced using the
bolometric flux measured  by NuSTAR. The best fit model flux is
estimated to $F_{0.1-100}= 1.93 \times
10^{-10}$\,erg\,s$^{-1}$\,cm$^{-2}$. According to our computations
the bolometric flux of this model is about 70\% of the total
luminosity, and the corresponding correction factor is $C\approx1.43$.
We used both the WD mass and the magnetospheric radius to derive the mass-accretion rate using the relation
\be
   C\times F_{0.1 -100} = \frac{GM\dot M}{4\pi D^2}\left(\frac{1}{R}-\frac{1}{R_{\rm m}}\right).
\ee
We obtained $\dot M \approx 1.3 \times
10^{15}$\,g\,s$^{-1}$  (or $\approx 2 \times
10^{-11}$\,M$_\odot$/yr). Here we employed the mass-radius
relation for WDs from \citet{Nbg:72}:
\be
       R = 7.8\times 10^8 \left[\left(\frac{1.44}{m}\right)^{2/3}-\left(\frac{m}{1.44}\right)^{2/3}\right]^{1/2} {\rm cm}.
       \label{eq:rm}
\ee
The uncertainties in $M$ and $R_{\rm m}$ almost cancel each
other, and only the uncertainty in the distance is important, which
is small in our case. The derived luminosity of EX Hya is thus $L
\approx 7.5\times 10^{31}$\,erg\,s$^{-1}$.

Formally, the evaluated mass-accretion rate is so low that EX~Hya,
having an orbital period of $P_{\rm orb} = 1.638$\,h, already enters the
region occupied by post-period-minimum CVs according to theoretical
expectations for the $\dot M - P_{\rm orb}$ relation \citep[see,
  e.g.][]{Howell.etal:01, GN:15}. On the other hand, the recently
measured secondary mass $M_2 = 0.10\pm 0.02$\,M$_\odot$
\citep{Echevarria.etal:16} excluded this possibility. Therefore, EX
Hya is just a system in a low-accretion state. Indeed, \citet{VKS:80}
(see also \citealt{Bateson:79}) remarked that EX Hya is a dwarf nova
with very short duration ($\approx\,4^{\rm d}$) and rare (every
574$^{\rm d}$) outbursts. Our result supports this statement.

Using the determined mass-accretion rate we can also evaluate the
magnetic field strength $B$ on the WD surface. Assuming that the
magnetospheric radius is proportional to the standard Alfv\'{e}n radius
$R_{\rm m} = \Lambda R_{\rm A}$, we get
\be
  r_{\rm m} = \frac{R_{\rm m}}{R} = \Lambda \left(\frac{B^4 R^5}{2GM\dot M^2}\right)^{1/7},
\ee
or
\be
      B = (2GM)^{1/4}\,\dot M^{1/2} R^{-5/4} \left(\frac{r_{\rm m}}{\Lambda}\right)^{7/4}.
\ee
Inserting numerical values, this implies $B=2.9\times 10^4$\,G
assuming that $\Lambda=0.5$. This low value is, however, slightly
higher in comparison with previous results \citep[see,
  e.g., ][]{Revnivtsev.etal:11}.

\subsection{NuSTAR observations of GK Per}

 \begin{figure} 
\begin{center}  
\includegraphics[width=  0.9\columnwidth]{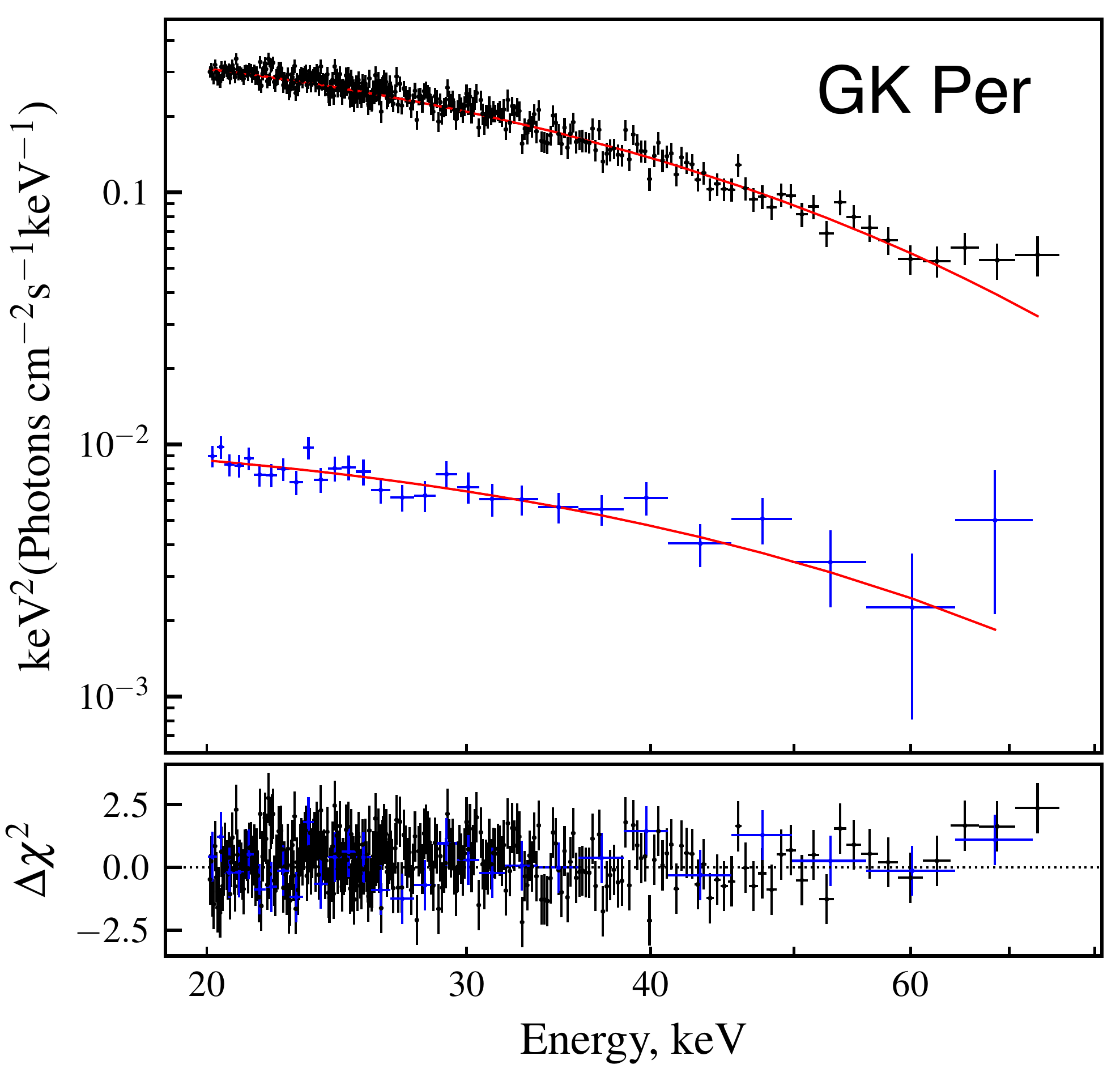}
\includegraphics[width=  0.9\columnwidth]{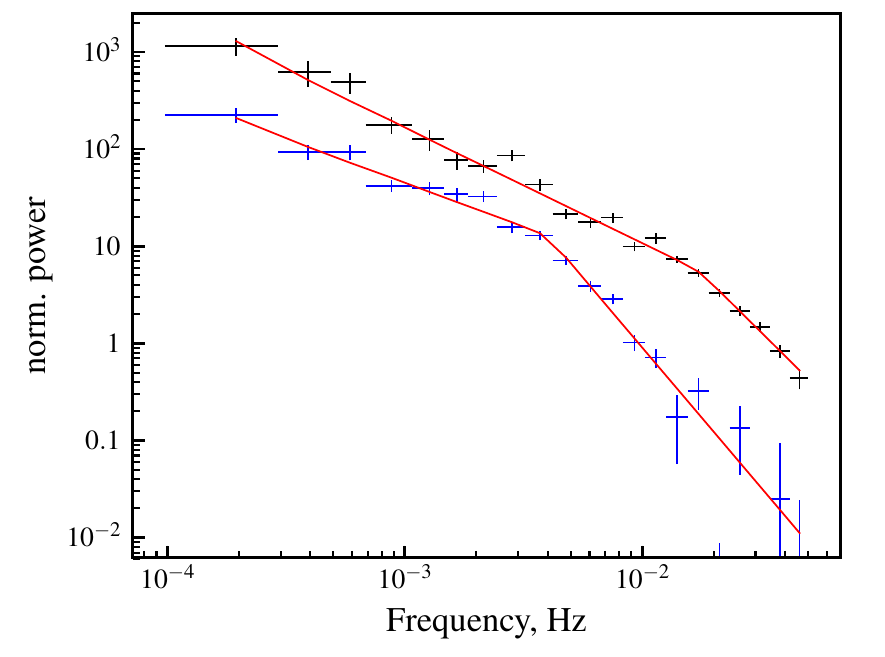}
\caption{\label{fig:gkper_sp} 
Top panel: NuSTAR spectra of GK Per during outburst and quiescence  fitted with short PSR models
 ($M = 0.79$\,M$_\odot$, $R_{\rm m} = 3.18\,R$ for outburst and $R_{\rm m} = 8.5\,R$ in quiescence).
 Spectra from both NuSTAR units combined for plotting to enhance clarity.  Bottom panel: 
The corresponding power spectra fitted with broken power laws, $\nu_{\rm br} = 0.017$ (outburst) and $\nu_{\rm br} = 0.004$
(quiescence). 
} 
\end{center} 
\end{figure}

 \begin{figure} 
\begin{center}  
\includegraphics[width=  0.9\columnwidth]{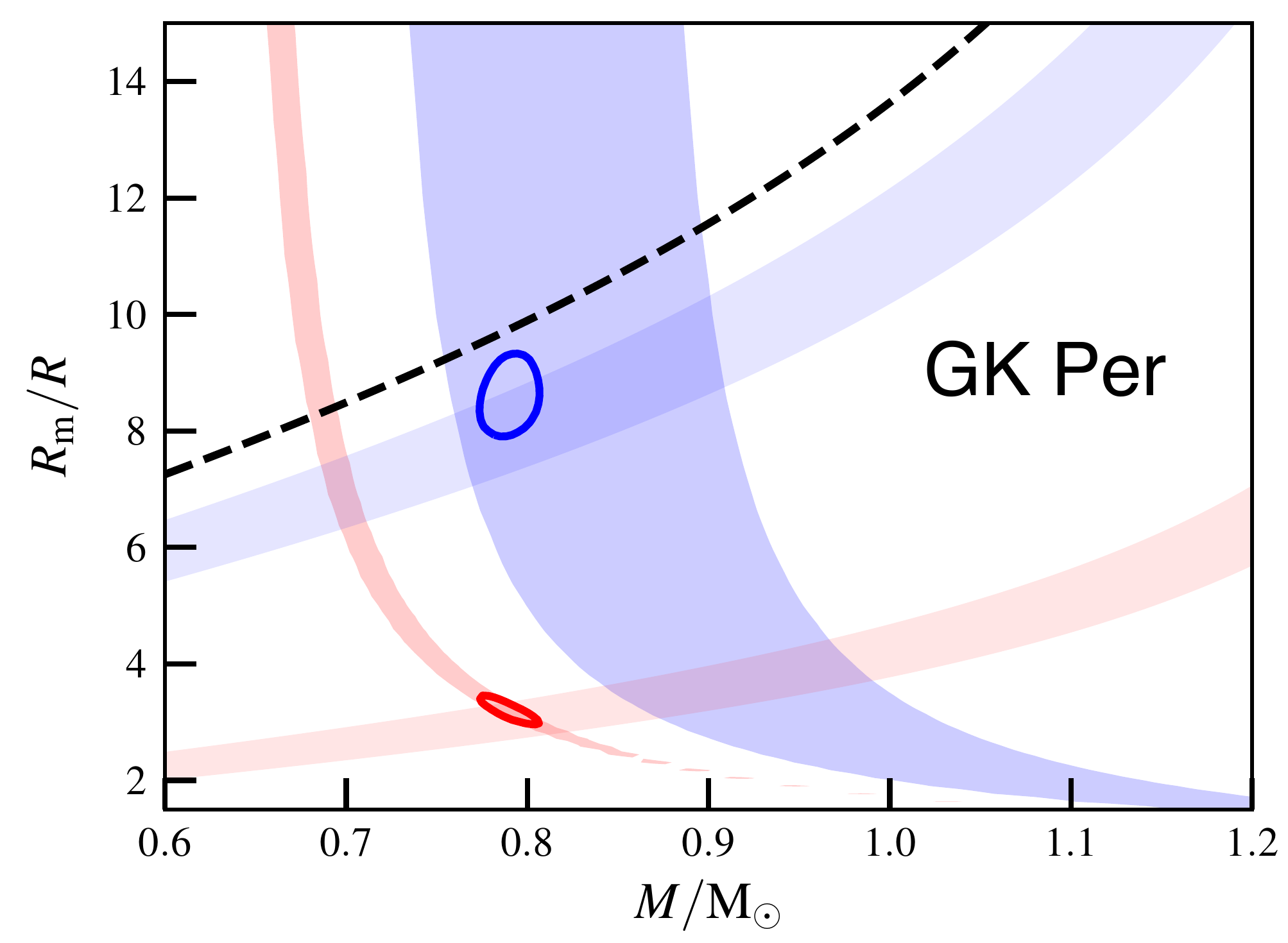}
\caption{\label{fig:gkper} 
Strips in the $m - r_{\rm m}$ plane obtained using spectral fitting of GK~Per
during outburst (red strips) and quiescence (blue strips) by short column PSR models, and using 
the break frequency in the power spectrum during outburst 
($\nu_{\rm br} = 0.017$\,Hz) and quiescence ($\nu_{\rm br} = 0.004$\,Hz). 
The corresponding most probable regions are shown by the solid closed curves.
The dashed curve corresponds to 
the corotation radius.
} 
\end{center} 
\end{figure}

The old nova GK Per is an IP which exhibits dwarf nova behavior as
well \citep{Watson.etal:85, Sim:02}. Therefore,  the states of the system with
 the very different accretion rates are observed, so the magnetospheric radius in this
system is expected to differ in quiescence and outburst. We
attempted to investigate whether this is indeed the case
\citep{Suleimanov.etal:16} using a NuSTAR observation of GK~Per during
the latest outburst \citep{Zemko16}, and \emph{Swift}/BAT and INTEGRAL hard X-ray spectra of
the source in quiescence. This allowed to estimate the magnetosphere size
respectively using observed break frequency, and from PSR model under assumption
that the mass of WD is the same in both cases. No dedicated observations
of the source quiescence were available at the time
\citep{Wada.etal:18} so we were not able to investigate the power spectrum. 
Note that \citet{Wada.etal:18} used a similar
approach to study the magnetospheric radius changes, but they employed
another spectral model, and obtained consistent results. In
particular, they derived a WD mass of $0.87\pm0.07$\,M$_\odot$ which
is in very good agreement with our result $0.86\pm0.02$\,M$_\odot$.

A NuSTAR observation of GK~Per in quiescence became available since
then, and this dataset allows us to directly measure the power
spectrum of the source. Its shape is similar to
that during the outburst, and the break is clearly detected.
The break occurs at
significantly lower frequency, as expected. This is illustrated in
Figs.\,\ref{fig:gkper_sp} and \ref{fig:gkper}. The best fits to the GK
Per spectra and power spectra during outburst and quiescence are shown
in Fig.\,\ref{fig:gkper_sp}. Again, we fit both power and energy
spectra simultaneously as described above for EX Hya, with
break frequencies in outburst and quiescence linked to the respective $R_m$
parameter values of the PSR model. The mass of the WD was linked for
both data sets. The results are presented in
Fig.\,\ref{fig:gkper}. The contours for the WD mass and magnetosphere
size for the two states obtained from the joint fit are shown by solid
curves. For illustration we also plot the constraints which are
obtained by independent fitting of energy and power spectra (as
colored strips in the plot). From the joint fit we obtain
$M=0.79\pm0.01$\,M$_\odot$ and $R_{\rm m}/R = 3.18\pm0.17$ (at
outburst) and $R_{\rm m}/R = 8.5\pm0.5$ (at quiescence). The obtained
WD mass and magnetospheric radii are slightly reduced compared to
our previous result. The reason for that is most likely related
to the fact that GK~Per is a transient system, so the mission-long
Swift/INTEGRAL spectra used previously to determine WD parameters
in queiscence contain also some outburst data.
We also note that the magnetospheric radius in
quiescence is close to the corotation radius.

The distance to GK Per is now constrained by Gaia DR2 to
$D=442\pm8.5$\,pc. This allows to estimate the mass-accretion rates
and corresponding luminosities during outburst ($7.87\times
10^{17}$\,g\,s$^{-1}$ and $5.8\times 10^{34}$\,erg\,s$^{-1}$)  and in
quiescence ($1.42\times 10^{16}$\,g\,s$^{-1}$ and $1.4\times
10^{33}$\,erg\,s$^{-1}$). The corresponding magnetic field strengths
on the WD surface are $B \approx 7.5\times 10^5$\,G and $B \approx
5.6\times 10^5$\,G. This difference indicates that the
magnetospheric radius dependence on accretion rate deviates slightly from one
expected from Alfv\'{e}n
model. Indeed, one would expect that the ratio of the
magnetospheric radius in quiescence $R^q_{\rm m}$ to the measured
one during outburst $R^o_{\rm m}$ has to be proportional to the mass-accretion
rates:
\be
           \frac{R^q_{\rm m}}{R^o_{\rm m}} \sim \left(\frac{\dot M^q}{\dot M^o}\right)^{-2/7} \approx 3.15\pm0.05.
\ee
The observed ratio is $2.7\pm0.21$, i.e., the discrepancy between the two
values is $\sim$\,2$\sigma$.  \citet{Wada.etal:18} obtained a higher ratio,
$3.9\pm0.5$, with systematically smaller measured magnetospheric
radii (based on spectral fits only).

\begin{figure*} 
\begin{center}  
\includegraphics[width= 0.6\columnwidth]{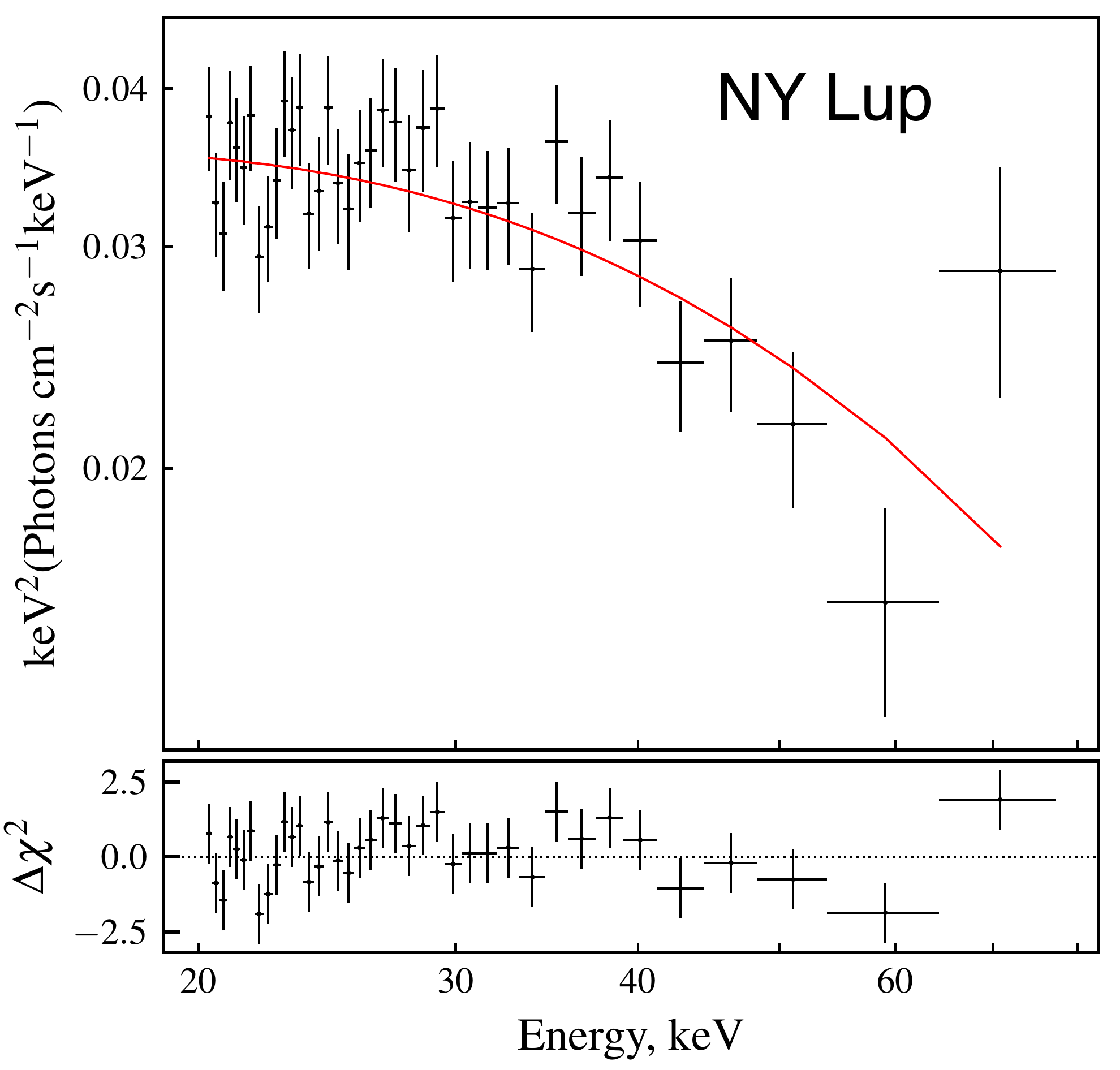}
\includegraphics[width= 0.6\columnwidth]{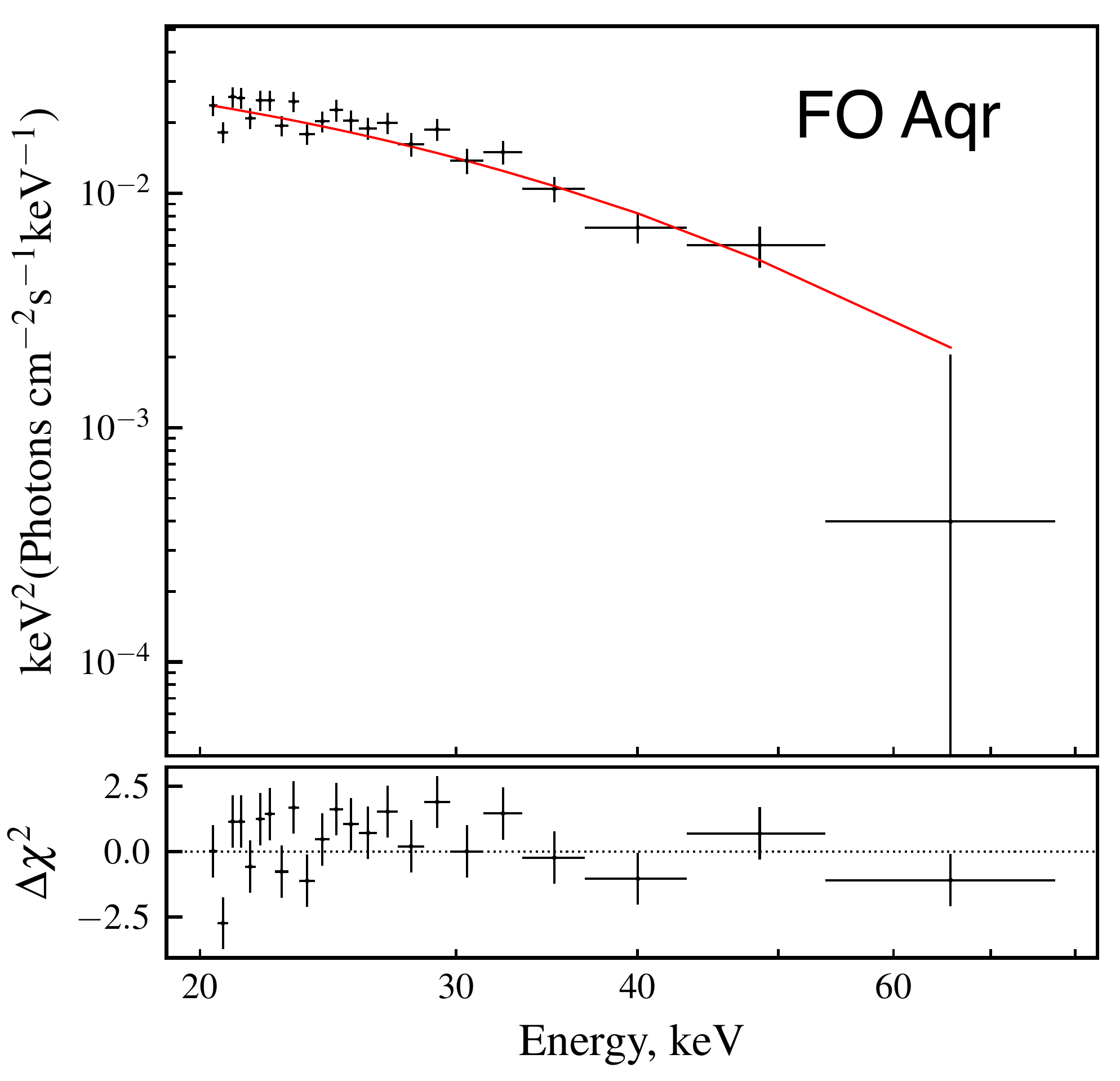}
\includegraphics[width= 0.6\columnwidth]{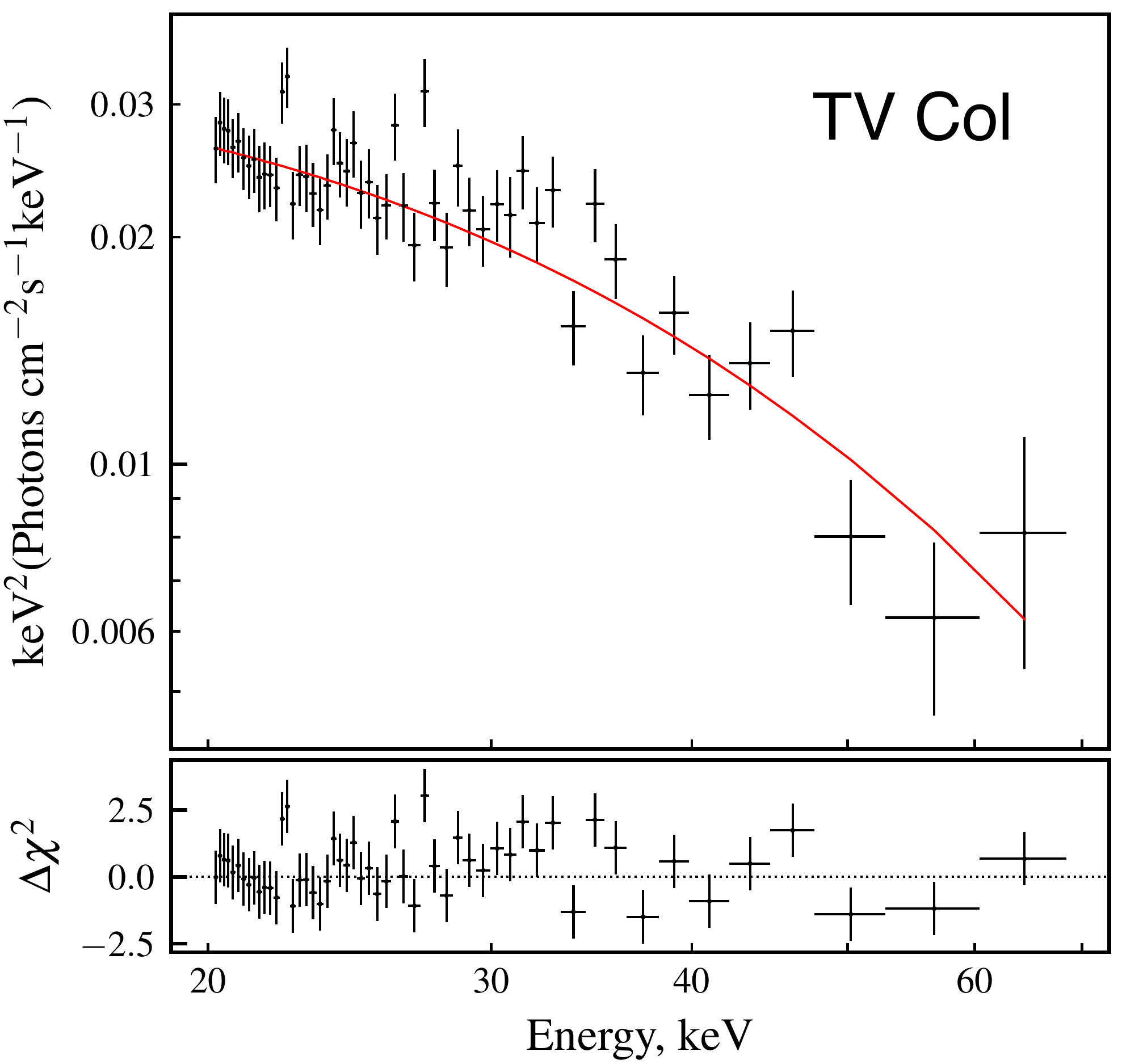}
\includegraphics[width= 0.6\columnwidth]{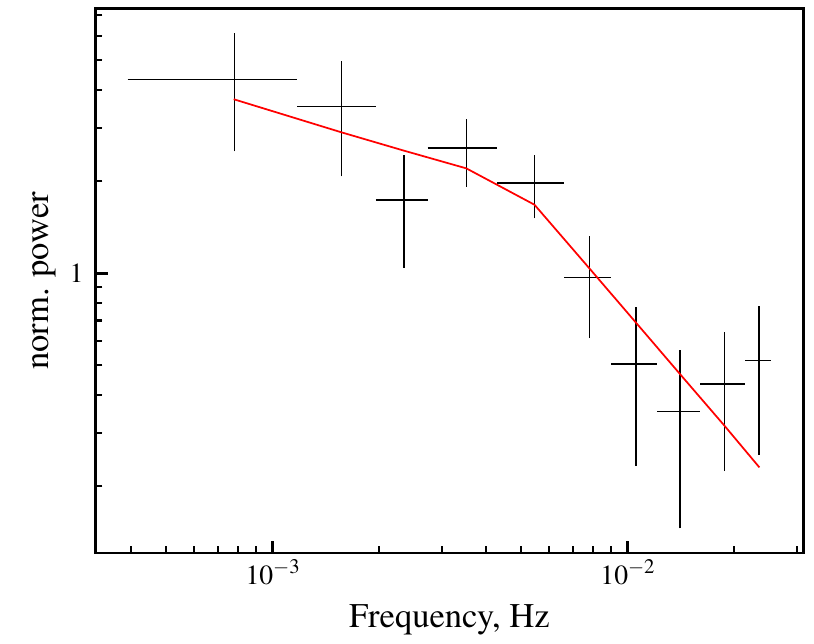}
\includegraphics[width= 0.6\columnwidth]{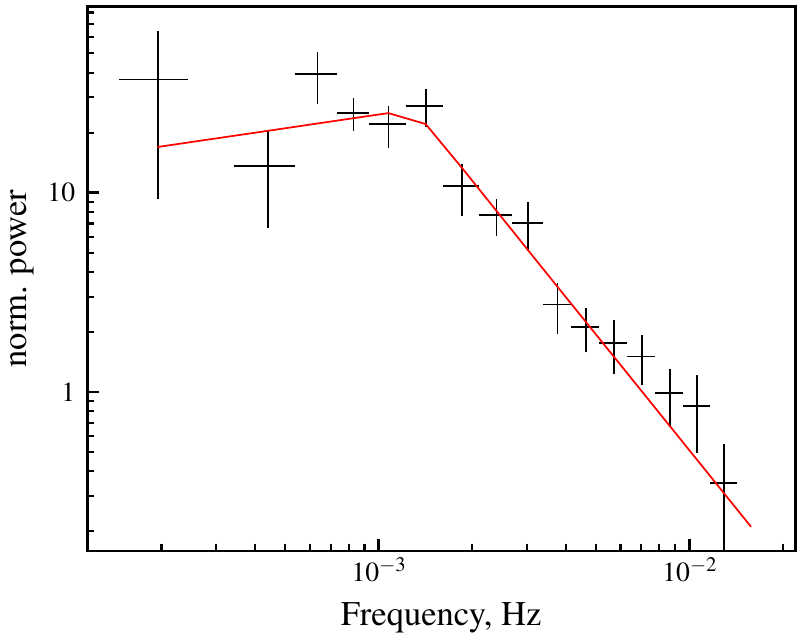}
\includegraphics[width= 0.6\columnwidth]{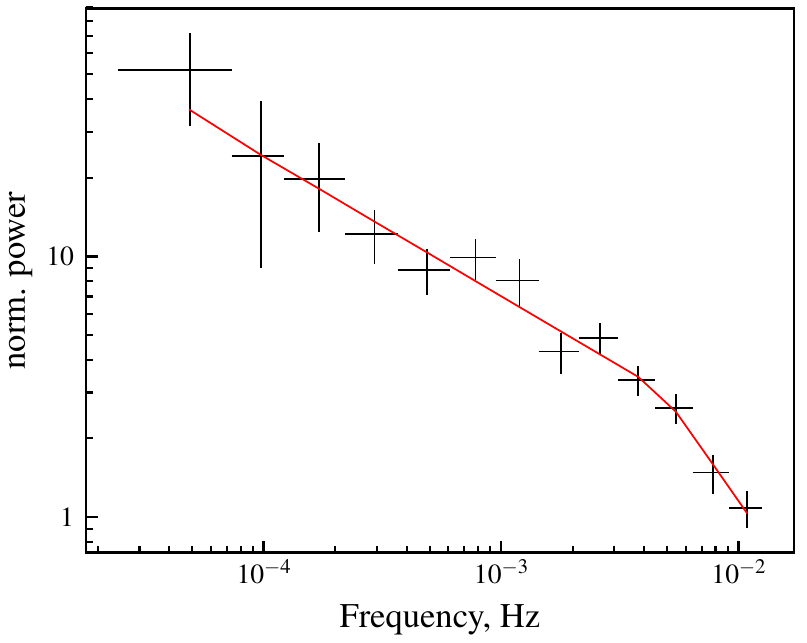}
\includegraphics[width= 0.6\columnwidth]{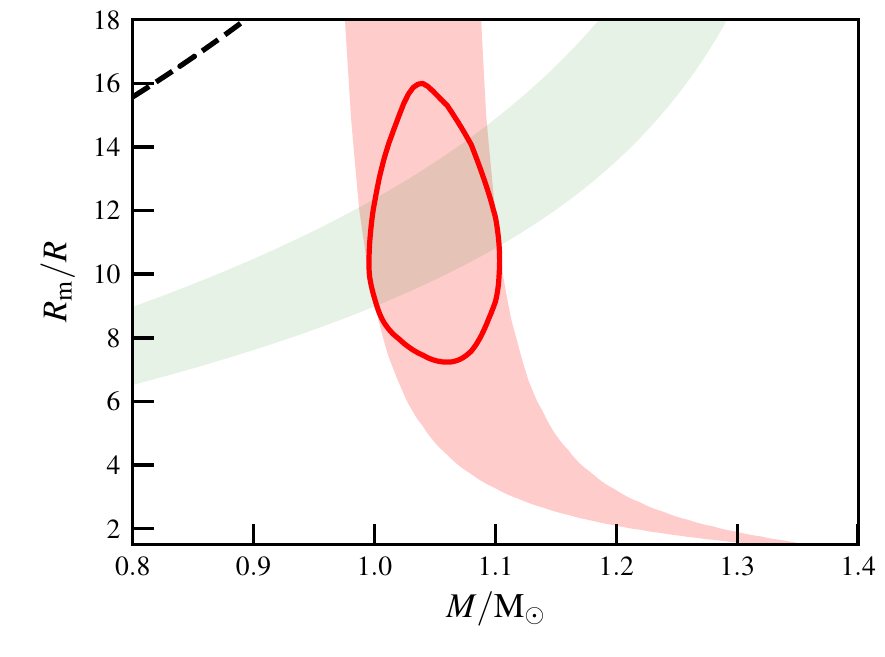}
\includegraphics[width= 0.6\columnwidth]{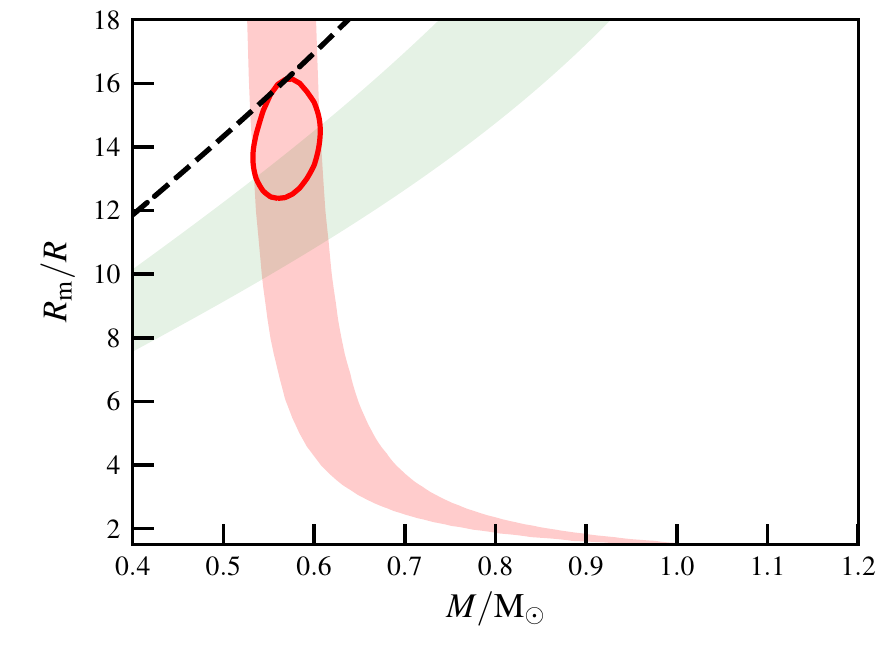}
\includegraphics[width= 0.6\columnwidth]{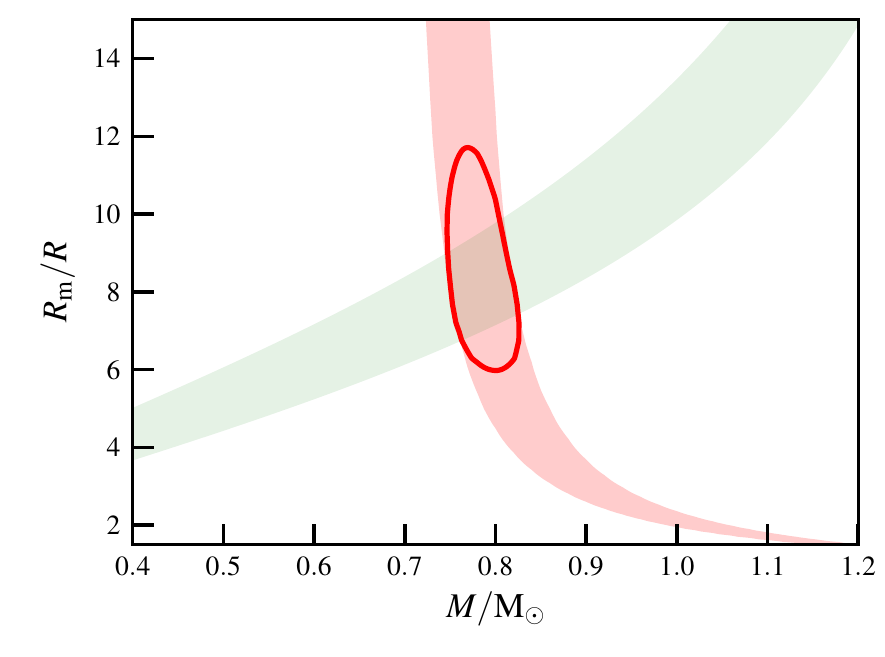}
\caption{\label{fig:other_all} 
Top panels: NuSTAR spectra of NY\,Lup, FO\,Aqr, and TV\,Col  fitted with short PSR models. 
Spectra from both NuSTAR units combined for plotting to enhance clarity. 
 Middle panels: 
The corresponding power spectra fitted with broken power laws.  Bottom panels: Strips in the $m - r_{\rm m}$ 
plane obtained using spectral fitting  (red strips) by short column PSR models, and using 
the break frequency in the power spectra (green strips).
The corresponding most probable regions are shown by the solid closed curves.
The dashed curves correspond to 
the corotation radii. The used fitting parameters are presented in Table\,\ref{tab1}.
} 
\end{center} 
\end{figure*}

\begin{figure*} 
\begin{center}  
\includegraphics[width= 0.4\columnwidth]{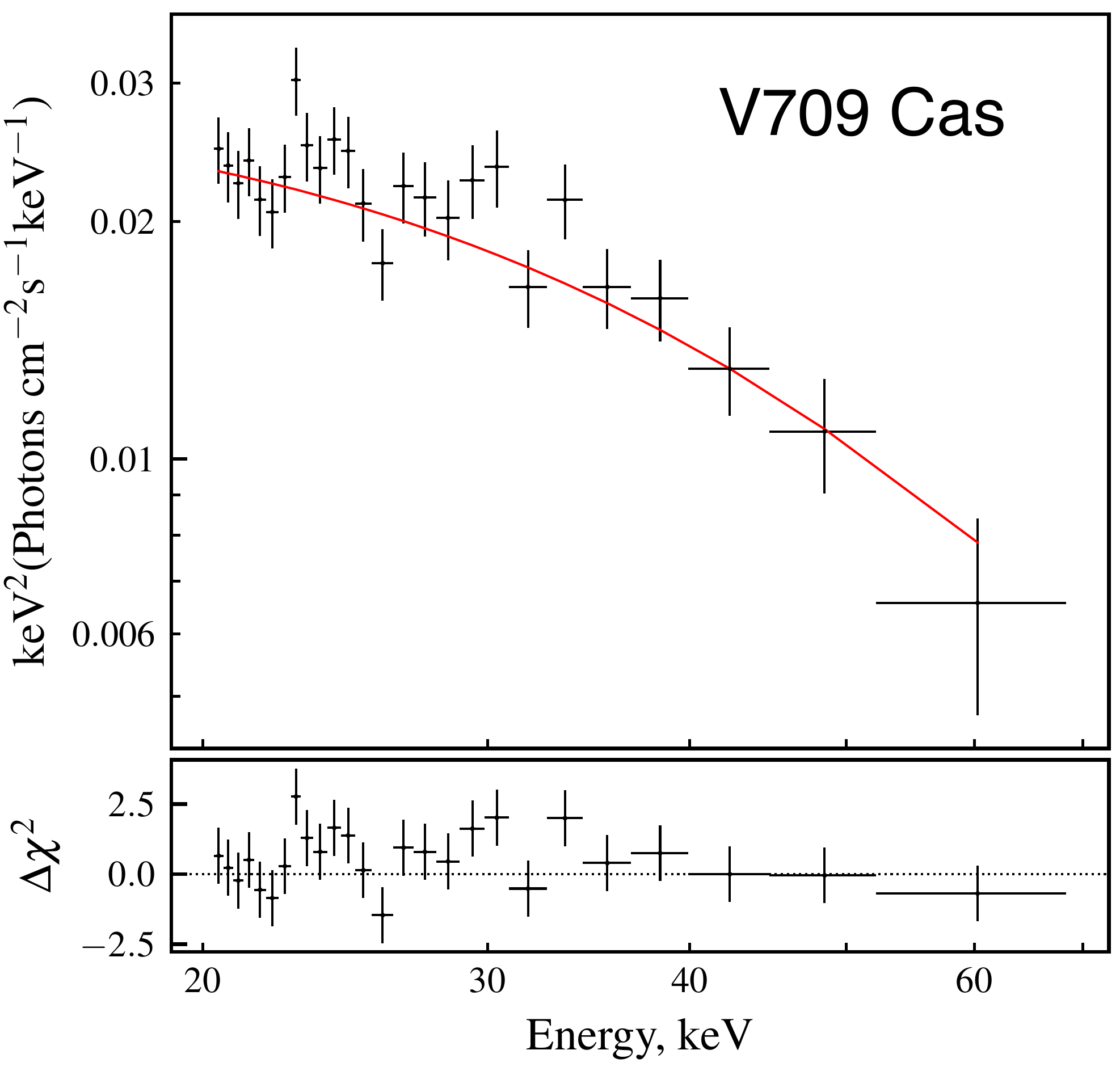}
\includegraphics[width= 0.4\columnwidth]{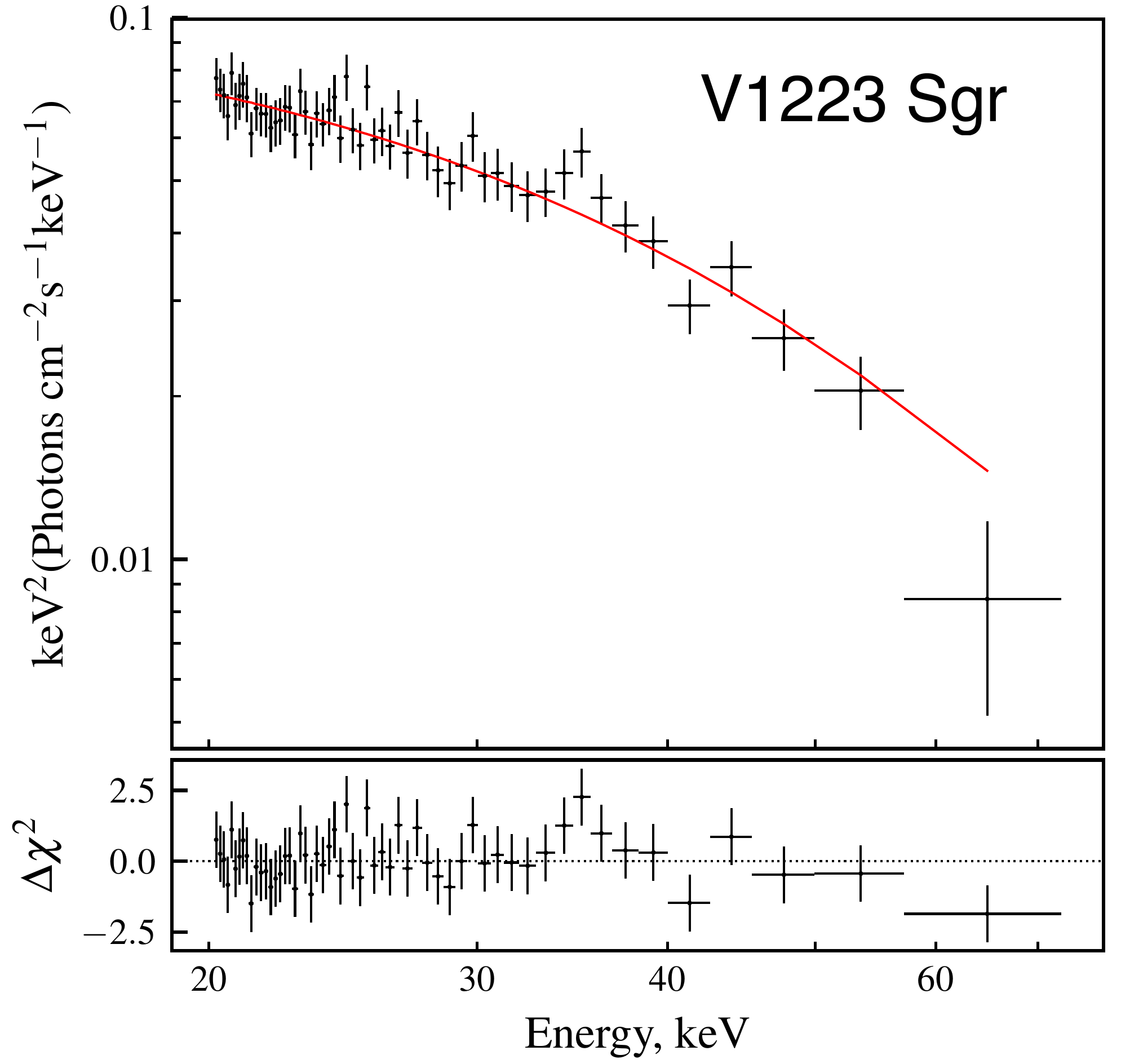}
\includegraphics[width= 0.4\columnwidth]{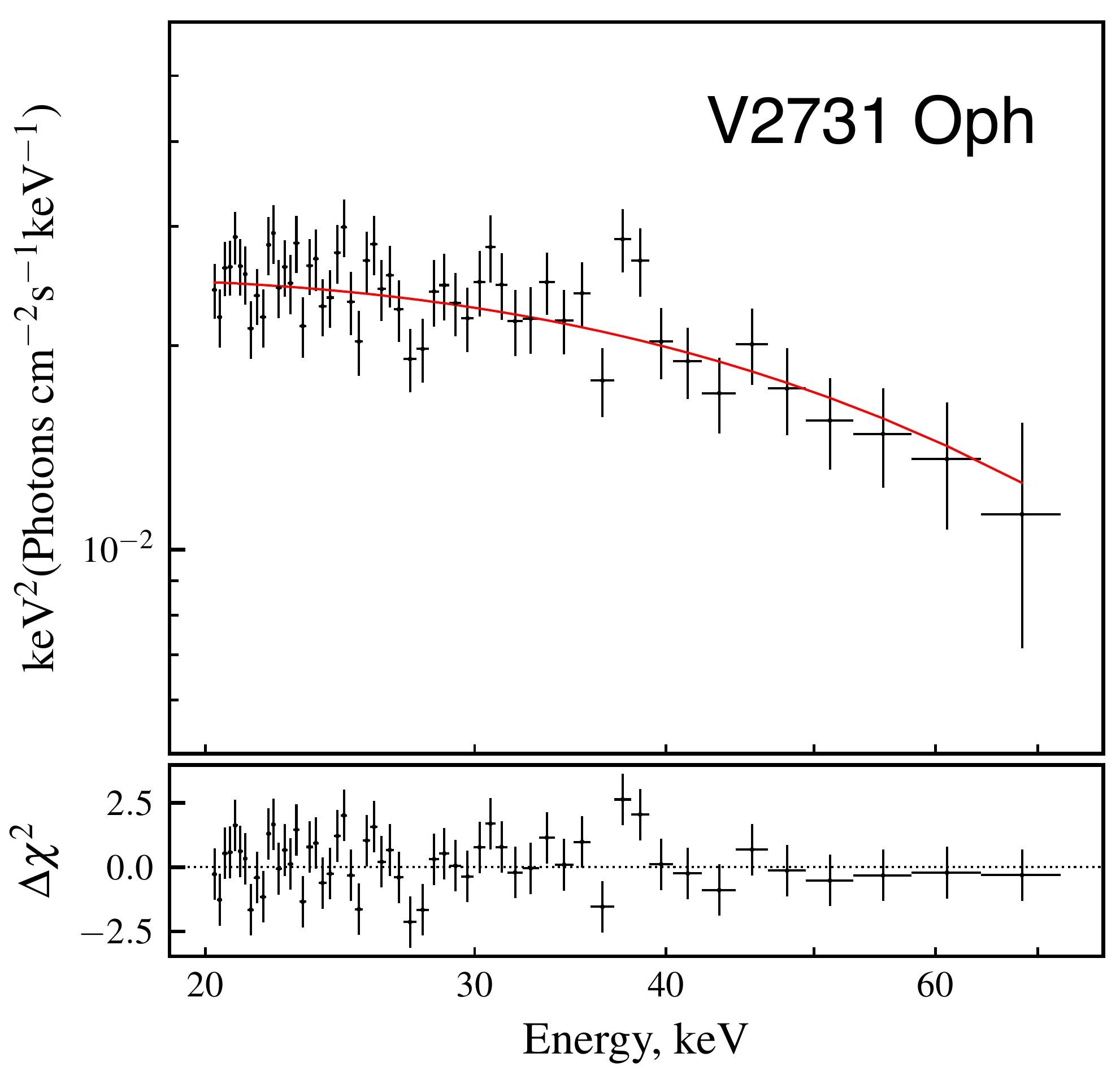}
\includegraphics[width= 0.4\columnwidth]{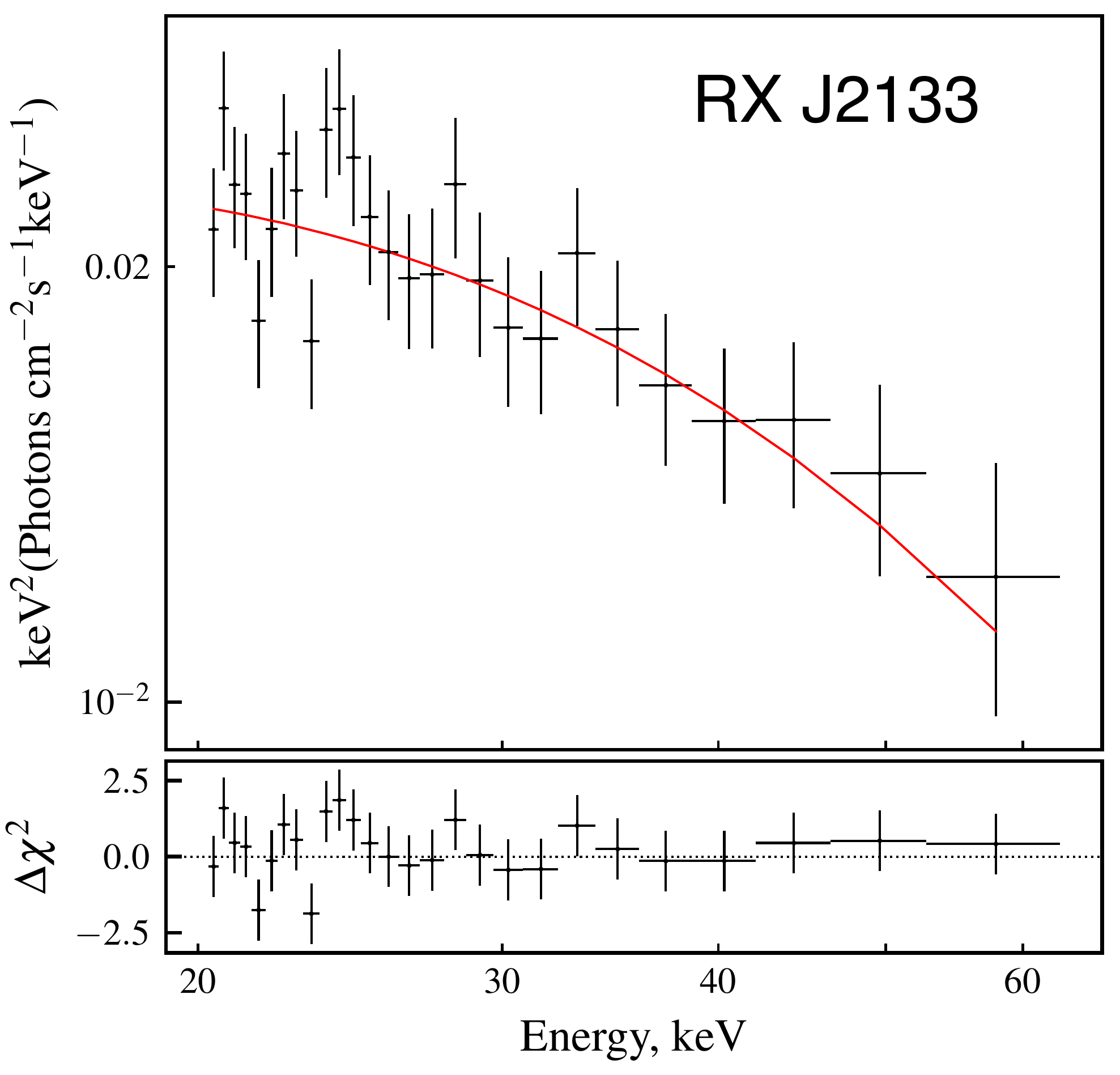}
\includegraphics[width= 0.4\columnwidth]{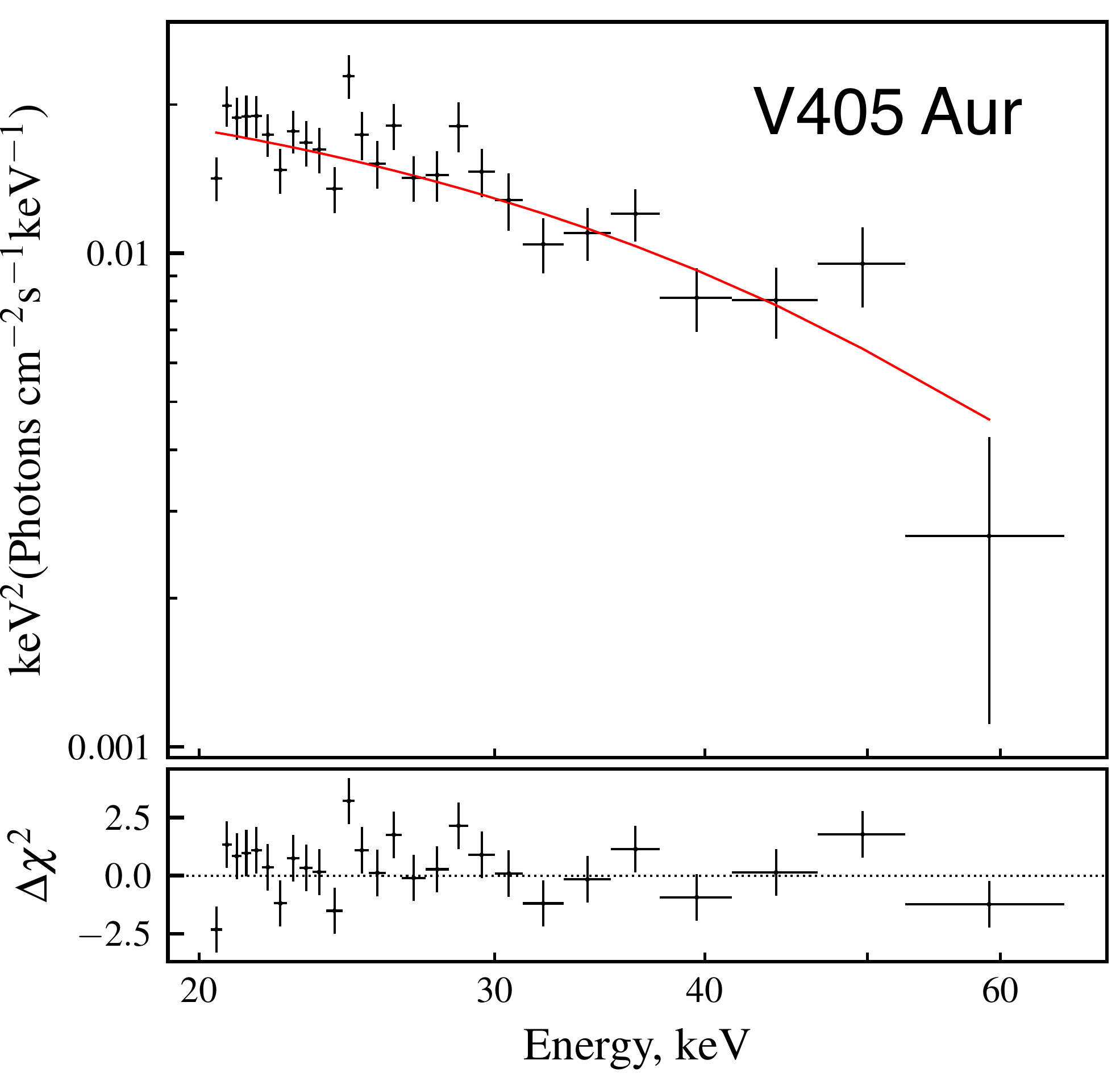}
\caption{\label{fig:other_sp} 
NuSTAR spectra of V709\,Cas, V1223\,Sgr, V2731\,Oph, RX\,J2133,  and V405\,Aur  fitted with short PSR models.  Spectra from both NuSTAR units combined for plotting to enhance clarity.
The used fitting parameters are presented in Table\,\ref{tab1}.
} 
\end{center} 
\end{figure*}

 \begin{table}
\centering   
\caption{Observed and derived parameters of IPs observed by NuSTAR. 
 \label{tab1} 
 }
\begin{tabular}[c]{l c c c c c }
\hline
Name & 	$M/$M$_\odot$        &  $\nu_{\rm br}$, Hz    & $R_{\rm m}/R$ & $F^b_{0.1-100}$        \\ 
\hline 
GK Per  &  $0.79 \pm   0.01$  &  0.017  & 3.18 $\pm$0.17&  24.79$\pm$0.5  \\
         &    &  0.004 & 8.5$^{+0.5}_{-0.4}$ &  0.59$\pm$0.03  \\
NY Lup &   $1.05 \pm  0.04$ &   0.005 &   10.3$^{+2.8}_{-2.0}$ &  2.08$\pm$0.11  \\
FO Aqr &  $0.57 \pm   0.03 $&  0.0013 &  14.1$\pm$1.1 &   2.38$\pm$0.31   \\
V2731 Oph & $  1.06 \pm   0.03 $&   &  7.8$^a$  & 1.47$\pm$0.07 \\
V709 Cas & $ 0.83 \pm 0.04 $&    & 9.6$^a$ &  1.57$\pm$0.14  \\
EX Hya &   $0.70 \pm   0.04$ &   0.013 &   3.4$^{+0.4}_{-0.3} $&  1.93$\pm$0.42  \\
V1223 Sgr & $ 0.72 \pm   0.02 $&   &   14.4$^a$ &   5.39$\pm$0.32  \\
V405 Aur &  $0.73\pm  0.03 $&  &   12.0$^a$ &  1.31$\pm$0.14 \\
J2133+5107 & $  0.95 \pm  0.04 $ &  &   17.2$^a$ &   1.35$\pm$0.09  \\
TV Col &  $ 0.79 \pm   0.03 $&   0.005  & 7.6$^{+2.6}_{-1.2}$ &   1.87$\pm$0.12  \\
 \hline
\end{tabular}
\begin{flushleft}{ 
Notes:  (a) Relative corotation radius; 
(b) Best fit model flux in the range 0.1--100\,keV in units $10^{-10}$\,erg\,s$^{-1}$\,cm$^{-2}$.
}\end{flushleft} 
\end{table}

\subsection{Other intermediate polars observed with NuSTAR}

Up to now, observations of ten IPs were performed by NuSTAR, and we
applied the approach described in detail above to all of them. The
available data does not always allow to obtain a power spectrum
with sufficient quality, particularly for slower rotating sources, and thus to detect the break. 
The
magnetospheric radii were estimated, therefore, using the break frequency
whenever possible, or fixed to the corotation radius otherwise. The
results are presented in Table\,\ref{tab1} and
Figs.\,\ref{fig:other_all} and \ref{fig:other_sp}.

Masses of three IPs (NY Lup, V1223 Sgr, and V709 Cas) were determined
by \citet{Shaw.etal:18} using NuSTAR observations. The reported values
are $1.16^{+0.04}_{-0.02} M_\odot$, $0.75\pm0.02 M_\odot$, and
$0.88^{+0.05}_{-0.04} M_\odot$, respectively. Comparison with our
estimates suggests that we obtain slightly smaller masses. There are
several possible reasons for this discrepancy.


First of all, we restrict our analysis to the energy range above
20\,keV due to the apparent complexity of the spectrum at lower
energies which is not accounted for by the model. While it is clear
that reflection or complex absorption must be responsible for the
observed deviations from PSR model, as already discussed above, there
is no physically motivated quantitative description of these
deviations, yet \citep[see, however,
][]{Hayashi.etal:18}. \citet{Shaw.etal:18} considered the full NuSTAR
energy range and concluded that reflection is essential to describe
the broadband spectrum. However, a rather simplified description for
the reflection component was necessarily used, which could bias their
estimates. Indeed, the overall fit quality and thus estimated
properties of the reflection component are dominated by the soft
band. On the other hand, the reflection bump peaks around 20--40\,keV
\citep{MZ:95}, i.e., close to the expected rollover of the PSR
spectrum. As a consequence, the rollover energy ultimately defining
the estimated WD mass, becomes strongly dependent on the ad-hoc
description of the soft part of the spectrum, which is clearly not
desirable. We thus restricted the analysis to the energy range which,
from a physical point of view, gives a real constraint for the
parameter of interest, i.e., the WD mass.

Note that \citet{Hailey.etal:16} also estimated WD masses for two
IPs observed with NuSTAR, namely TV~Col and V2731~Oph. They used an
approach similar to ours and for the same reasons restricted their analysis
to the $\ > 15$\,keV energy range. As a result, they found masses of
$0.77\pm 0.03$\,M$_\odot$ and $1.16\pm0.05$\,M$_\odot$ respectively,
which are close to our estimates for these sources.

It appears also that the difference between our results and those
obtained by other authors is more significant for heaviest
WDs. We believe that the reason for this is the use of the different
models of the PSR. \citet{Hailey.etal:16} and \citet{Shaw.etal:18}
used the models computed by \citet{SRR:05} in cylindrical geometry for
a fixed local mass-accretion rate $a= 1$\,g\,s$^{-1}$. This value is
comparatively low and implies a tall PSR for heavy WDs. As a result,
the maximum PSR temperature in such models is lower than the maximum
temperature in the case of a short PSR, and a more massive WD is
required to reproduce the observed spectrum. On the other hand, our
model spectra \citep{Suleimanov.etal:16} are computed for a fixed
$\dot M= 10^{16}$\,g\,s$^{-1}$ and relative PSR footprint area $f =
5\times 10^{-4}$. Under these conditions the local mass-accretion rate
increases with the WD mass so that the PSRs remain short for any WD
mass, i.e., our model is self-consistent. For low-luminosity IPs,
where the PSR might indeed be tall, a separate model grid must be used
to avoid a bias in the WD mass determination.

\section{Swift/BAT observations} 
 
While NuSTAR provides data of exceptional quality, the number of
dedicated observations is limited. To increase the sample of objects,
we used, therefore, also the energy spectra accumulated throughout the
Swift mission lifetime, as mentioned above. The Swift/BAT spectra
cover the energy range above 15\,keV, so the results shall be
comparable with those obtained with NuSTAR. We have identified 35
sources significantly detected in the Swift/BAT Survey and known as
IPs\footnote{https://asd.gsfc.nasa.gov/Koji.Mukai/iphome/catalog/alpha.html}. The
final list of objects can be found in Table\,\ref{tab2}. As no timing
is available for most objects, we assumed corotation ($R_{\rm m} =
R_{\rm c}$) in these cases. We considered also both, short and tall
PSR models for all objects, and list the fit results for the tall PSR
where observed flux and estimated distance indicate this model might
be better justified. 

\begin{figure} 
\begin{center}  
\includegraphics[width=  0.9\columnwidth]{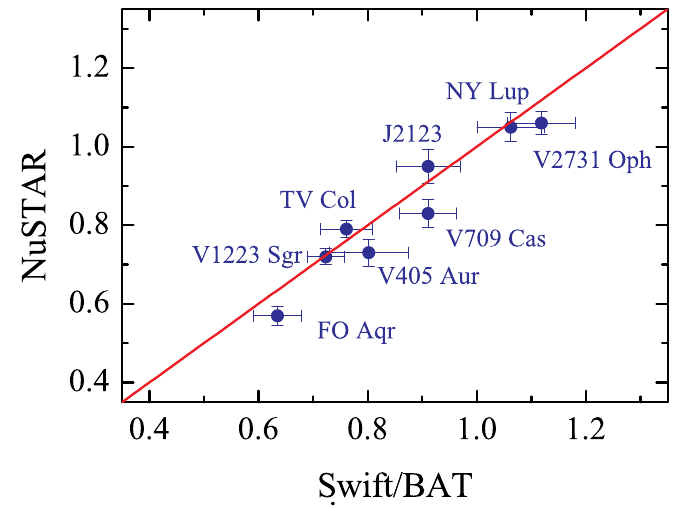}
\caption{\label{fig:SwNu} 
Comparison of the WD masses obtained by fitting their spectra observed by NuSTAR and Swift/BAT.
} 
\end{center} 
\end{figure}

Note that we excluded GK~Per and EX~Hya from the comparison
because the assumption that the matter falls from the corotation radius is
clearly inapplicable for these objects. As illustrated in
Fig.\,\ref{fig:SwNu}, the results obtained for NuSTAR and Swift/BAT
are otherwise broadly compatible for the objects observed by both
observatories, as expected. Inclusion of the Swift/BAT data allows,
however, to significantly increase the source sample, to compare our
results with previous investigations, and to assess how various
theoretical uncertainties might affect the deduced WD masses for the
entire population.

 \begin{figure} 
\begin{center}  
\includegraphics[width=  0.9\columnwidth]{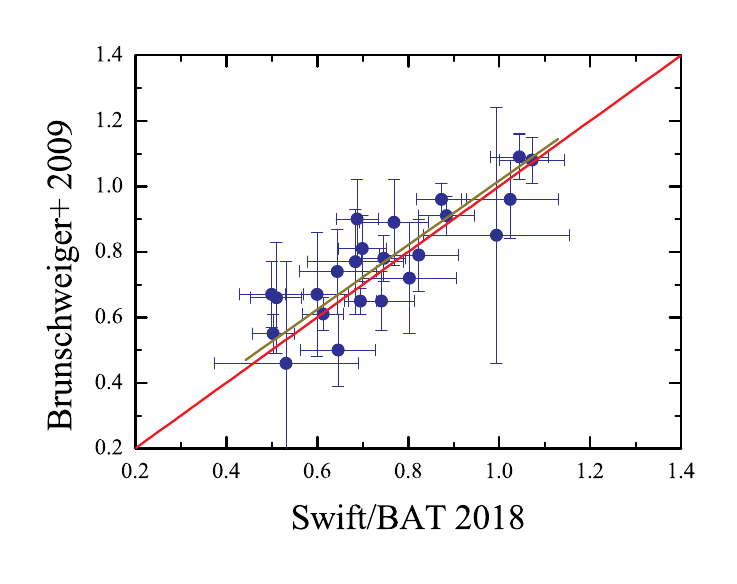}
\caption{\label{fig:SwBr} 
Comparison of the WD masses obtained by \citet{Betal:09} and by us. 
} 
\end{center} 
\end{figure}

 \begin{figure} 
\begin{center}  
\includegraphics[width=  0.9\columnwidth]{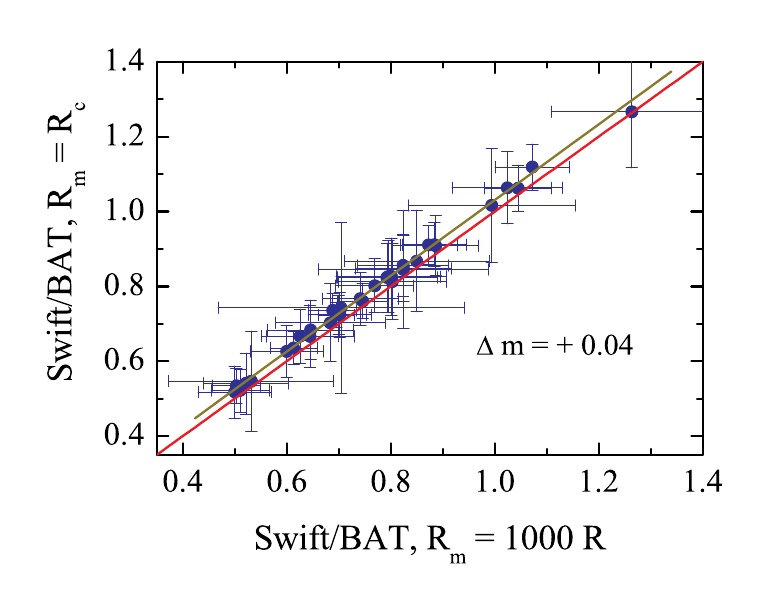}
\caption{\label{fig:Swcor} 
Importance of the finite magnetospheric radii for WD mass determination.
} 
\end{center} 
\end{figure}

 \begin{figure} 
\begin{center}  
\includegraphics[width=  0.9\columnwidth]{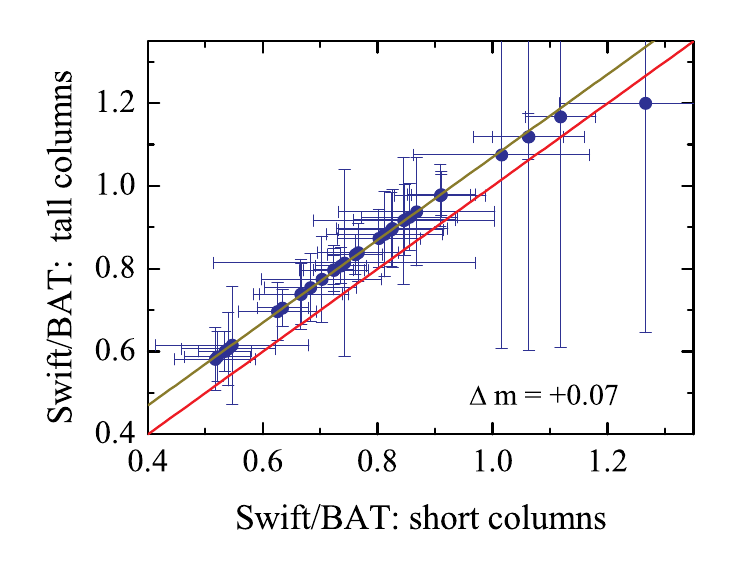}
\caption{\label{fig:Swclmn} 
Systematic difference between WD masses derived with using short and tall PSR models.
} 
\end{center} 
\end{figure}

 \begin{figure} 
\begin{center}  
\includegraphics[width=  0.9\columnwidth]{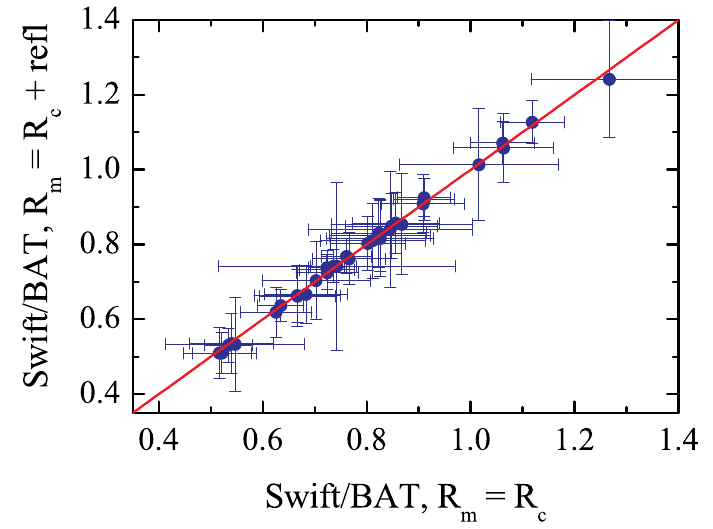}
\caption{\label{fig:Swrefl} 
Difference between WD masses derived using short PSR models with and without reflection component
taken into account.
} 
\end{center} 
\end{figure}

 \begin{figure} 
\begin{center}  
\includegraphics[width=  0.9\columnwidth]{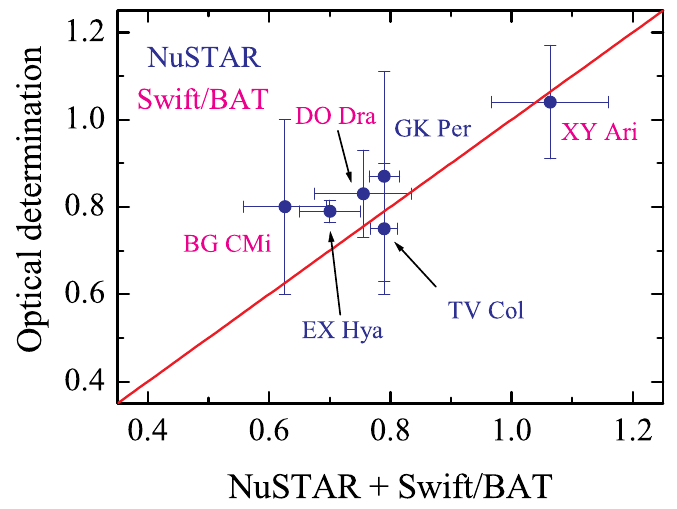}
\caption{\label{fig:opt} 
Comparison of the WD masses derived from X-ray spectra and obtained by optical methods.
} 
\end{center} 
\end{figure}

 \begin{figure} 
\begin{center}  
\includegraphics[width=  0.9\columnwidth]{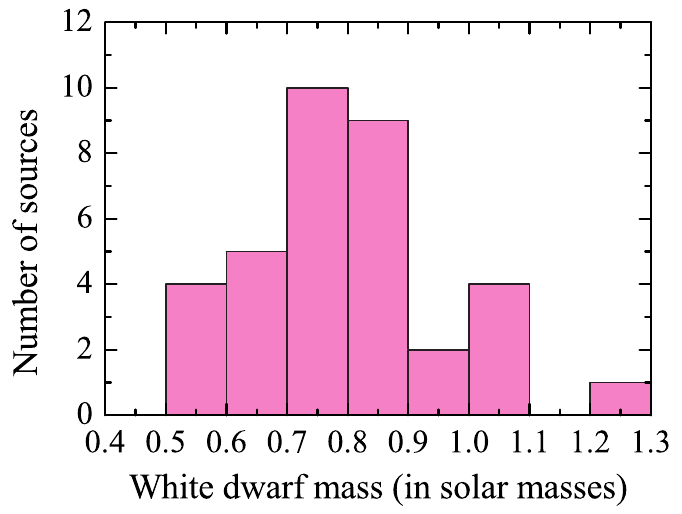}
\caption{\label{fig:dstr} 
Distribution of WD numbers over their masses.
} 
\end{center} 
\end{figure}

 \begin{figure} 
\begin{center}  
\includegraphics[width= 0.9\columnwidth]{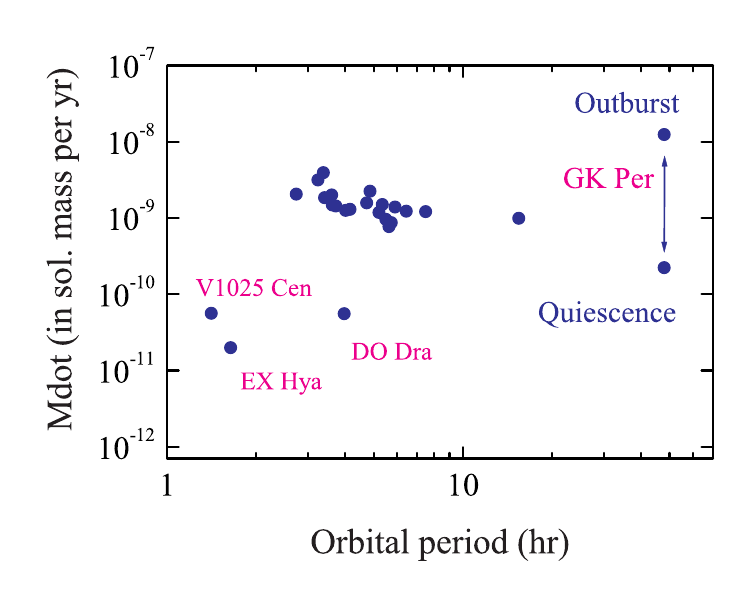}
\includegraphics[width= 0.9\columnwidth]{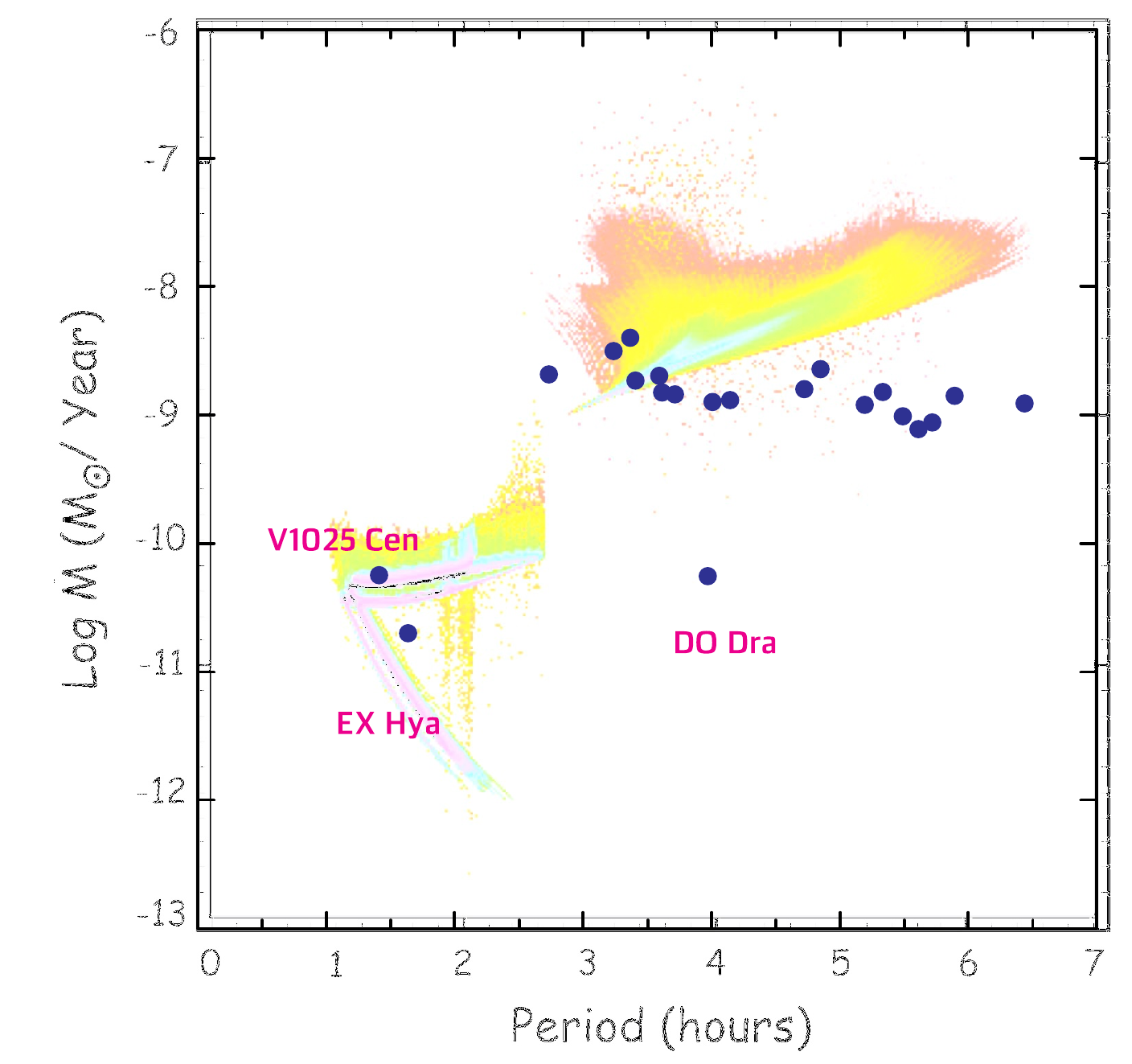}
\caption{\label{fig:mdot} 
Distribution of mass-accretion rates in the studied IPs vs.
orbital periods. Top panel: Only IPs with Gaia DR2 distances $<$1\,kpc
are shown. Bottom panel: Same results together with model
predictions by \citet{Howell.etal:01}; \copyright\ AAS; reproduced with
permission.
} 
\end{center} 
\end{figure}

 \begin{figure} 
\begin{center}  
\includegraphics[width=  0.9\columnwidth]{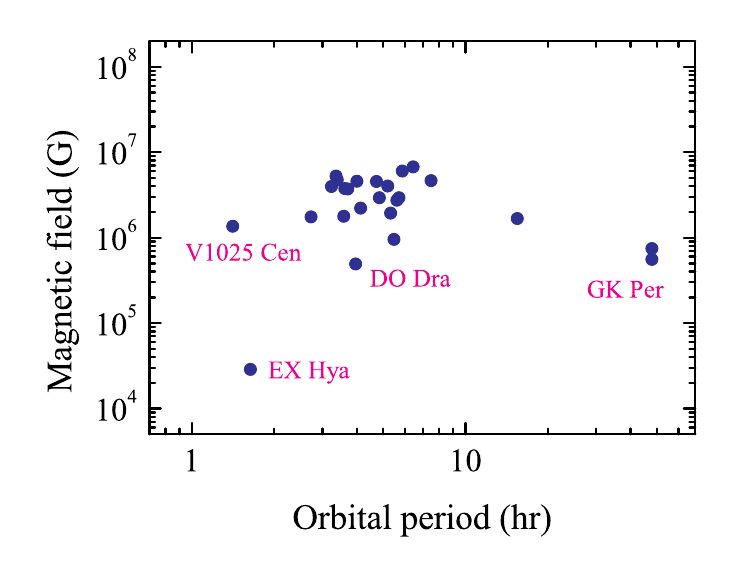}
\caption{\label{fig:magn}
Distribution of WD magnetic field strengths in the studied IPs
vs. orbital periods. Only IPs with the Gaia DR2 distances
$<$1\,kpc are shown.
} 
\end{center} 
\end{figure}

\section{Intermediate polar statistics}
\subsection{Comparison with published works and consistency checks}

First of all, we compare WD masses obtained by \citet{Betal:09} (23
IPs) with those found by us for the same sources and using the same
assumptions (Fig.\,\ref{fig:SwBr}). In particular, for this comparison
we used the short PSR models with the fixed relative magnetospheric
radius $R_{\rm m} = 1000\,R$ because \citet{Betal:09} used our old
models \citep{SRR:05} computed in cylindrical geometry assuming matter
falling from infinity. Fig.\,\ref{fig:SwBr} demonstrates that under
this approximation the difference between the old model grid and the
new one with a more realistic PSR cross-section is minor, and the
results are consistent within uncertainties.

We can now move one step further, and repeat the analysis using the
same model grid, but fixing the relative magnetospheric radii to the
corotation ones, which is more realistic. The comparison of the
obtained WD masses with those found using the assumption about matter
falling from  infinity is presented in Fig.\,\ref{fig:Swcor}. The new
masses are shifted to larger values, and the average displacement for
the sample is about $+0.04 M_{\odot}$, which is comparable with the
typical uncertainty in the WD mass determination. The conclusion one
could make is that on average IPs are sufficiently close to corotation
to make the resulting difference in spectral shape for the two
assumptions undetectable in available data. Note, however, that this
might not hold for dedicated pointed observations.

Another source of uncertainty is the assumed PSR height. We repeat the
analysis for all 35 sources using the new tall PSR model, again
assuming corotation. As expected, the results are shifted to higher WD
masses and the shift amounts to $+0.07 M_\odot$ (Fig.\,\ref{fig:Swclmn}). The
statistical uncertainties of the massive WDs measured using the new
grid are large, but this is just a consequence of the lower upper
limit ($1.2 M_\odot$) of computed WD masses in the tall PSR model
grid. This is most noticeable for the massive IP IGR\,J08390$-$4833.

We also investigated how the reflection component might bias our mass
estimates if only the hard X-ray band is considered. In particular, we
repeated the above analysis assuming a short PSR, but including a
reflection component {\tt reflect} with the maximum possible
reflection parameter $Refl=\Omega/2\pi=1$ (here $\Omega$ is the solid
angle of the reflecting slab as seen from the X-ray source). As
illustrated in Fig.\,\ref{fig:Swrefl}, the presence of reflection does
not affect the deduced masses. This confirms that our approach of
relying on hard X-ray data only is indeed justified.

\subsection{Statistical properties of the IP sample}

Now let us discuss the statistical properties of the entire sample.
We have determined the WD mass for a large number of IPs using the
same model and employing an individual approach to some better
investigated sources (see previous sections). The WD masses obtained
using NuSTAR observations have higher accuracy, and also give a better
understanding of the actual magnetosphere size, so we prefer them whenever
available (see Table\,\ref{tab1}). For the rest of the objects we use the
most recent Swift/BAT results. Consequently, we have now a uniform
sample of WD masses in 35 IPs. The final results for all objects
are summarized in Table\,\ref{tab2}. The uncertainties in the mass
accretion rates $\dot M$, the luminosities $L$, and the magnetic field
strength $B$ were estimated taking only the uncertainties in distance
(the largest source of uncertainty) into account.

\begin{table*}
\centering   
\caption{Observed and derived parameters of the investigated IPs.
 \label{tab2} 
 }
\begin{tabular}[c]{l c r r r r c c c c }
\hline
Name & 	$M/$M$_\odot$        &  $P_{\rm orb}$, hr  & $P_{\rm spin}$, s & $R_{\rm m}/R$ & $D$, pc & $F^b_{0.1-100}$ & $\dot M$, 10$^{16}$g/s & $L$, 10$^{33}$\,erg/s & $B$, MG   \\ 
\hline 
EX Hya$^*$ 	& 0.70$\pm$0.04 	&1.63762 & 4021.62 & 3.4 & 56.95$\pm$0.13&1.93 &  (13$\pm$0.6)$\times 10^{-2}$ &  (75$\pm$0.4)$\times 10^{-3}$ & (29$\pm$0.07)$\times 10^{-3}$	\\
TV Col 		& 0.79$\pm$0.03	 &5.4864 	&1911	 &7.6 		& 512.6$^{+4.7}_{-4.6}$		&	1.87  & 6.2$\pm$0.1 & 5.9$\pm$0.1 &	 0.96$\pm$0.01 \\
DO Dra$^*$ 		&0.76$\pm$0.08 	&3.96898 &529.3 	 &12.2$^a$ 	& 197.8$\pm$1.2 		&0.70 & 0.35 & 0.33$\pm$0.01 & 0.49$\pm$0.003 \\
V405 Aur 		&0.73$\pm$0.03 	&4.14288 &545.5 	&12.0$^a$ 	& 675$\pm$14 		& 1.31& 8.25$^{+0.35}_{-0.33}$ &	7.13$^{+0.30}_{-0.28}$ & 2.2$\pm$0.05 \\
MU Cam 		&0.67$\pm$0.08 	&4.7245 	&1187.22 &18.2$^a$ 	& 982$^{+28}_{-26}$ 	&0.66&10.06$^{+0.58}_{-0.53}$ & 7.56$^{+0.44}_{-0.40}$ & 4.6$\pm$0.13 \\
V667 Pup 	&0.86$\pm$0.08 	& 5.6112 	&512.42 	&13.9$^a$ 	& 814$^{+52}_{-46}$	&0.75 & 4.89$^{+0.65}_{-0.54}$ & 5.95$^{+0.79}_{-0.66}$ &  2.7$\pm$0.2 \\
V2400 Oph 	&0.72$\pm$0.05 	&3.408 	& 927.6 	&16.8$^a$	 & 715$\pm$17 		&1.68 & 11.7$^{+0.6}_{-0.5}$ &  10.26$\pm$0.5 &	 4.6$\pm$0.1 \\
BG CMi		&0.63$\pm$0.07 	& 3.23395&847.03	&14.3$^a$	&993$^{+57}_{-51}$	&1.12 &20.1$^{+2.39}_{-2.02}$	&13.2$^{+1.6}_{-1.3}$	&4.0$\pm$0.2 \\
V709 Cas	 	&0.83$\pm$0.04 & 5.3329	&312.75	&9.6$^a$		&747$\pm$12			&1.57&9.6$\pm$0.3	        	&10.5$\pm$0.3		&1.9$\pm$0.03	\\
V2731 Oph	&1.06$\pm$0.03 &15.42	       &128.1	&7.8$^a$	&832$^{+270}_{-164}$  	&1.47&6.3$^{+4.8}_{-2.3}$	&12.2$^{+9.2}_{-4.3}$	&1.7$^{+0.5}_{-0.3}$ \\
PQ Gem		&0.77$\pm$0.07 & 5.19262		&833.42	&16.7$^a$	&766$^{+22}_{-21}$	&1.06&7.6$\pm$0.4			&7.4$\pm$0.4			&4.1$\pm$0.1 \\
V1223 Sgr	&0.72$\pm$0.02 & 3.36586	&746	&14.4$^a$	&580$\pm$16			&5.39&25.1$^{+1.4}_{-1.3}$	&21.7$^{+1.3}_{-1.2}$	&5.3$\pm$0.15 \\
AO Psc		&0.53$\pm$0.05 & 3.59102	&805.2	&11.3$^a$	&495$^{+11}_{-10}$	&1.06&12.8$^{+0.6}_{-0.5}$	&6.1$\pm$0.3			&1.8$\pm$0.04 \\
V1025 Cen$^*$	&0.61$\pm$0.14 & 1.41024	&2146.6	&24.8$^a$	&193.2$^{+4.9}_{-4.6}$ 	&0.52&0.36$\pm$0.02  		&0.23$\pm$0.01		&1.4$\pm$0.03 \\
V1062 Tau	&0.81$\pm$0.10 & 9.98222	&3780	&49.1$^a$	&1582$^{+220}_{-172}$	&0.69&17.9$^{+5.3}_{-3.4}$	&20.7$^{+6.1}_{-4.3}$	&45$^{+6}_{-5}$ \\
XY Ari		&1.06$\pm$0.10 & 6.06473	&206.3	&10.7$^a$	&2000$^c$				&0.60&14.2				&28.6				&	4.5	\\
FO Aqr		&0.57$\pm$0.03 &4.84944	&1254.3	&14.1		&526$\pm$14			&2.38&14.3$^{+0.8}_{-0.7}$	&7.9$\pm$0.4			&3.0$\pm$0.1 \\
TX Col		&0.52$\pm$0.07 & 5.7192	&1911	&19.5$^a$	&512.6$^{+4.7}_{-4.6}$	&0.83&5.5$\pm$0.10		&2.6$\pm$0.05		&3.0$\pm$0.03 \\
GK Per		&0.79$\pm$0.01 & 47.9233	&351.3	&3.18		&442.0$^{+8.6}_{-8.3}$ 	&24.8&78.7$^{+3.1}_{-2.9}$	&57.9$^{+2.3}_{-2.2}$	&0.75$\pm$0.1 \\	
		&			    &	 	 	& 		&8.5	    	&							&0.59&1.42$^{+0.06}_{-0.05}$	&1.38$\pm$0.05		&0.56$\pm$0.1 \\
V2306 Cyg	&0.70$\pm$0.10 & 4.35708	&1466.7	&22.1$^a$	&1360$^{+65}_{-59}$	&0.49&13.0$^{+1.3}_{-1.1}$	&10.9$^{+1.1}_{-0.9}$	&7.8$\pm$0.4 \\
NY Lup		&1.05$\pm$0.04 & 9.864	&693.01	&10.3		&1272$^{+46}_{-43}$	&2.08 &20.7$^{+1.5}_{-1.4}$	&40.3$^{+2.3}_{-2.7}$	&4.9$\pm$0.2 \\ 
V1033 Cas	&1.02$\pm$0.15 & 4.032	&563.5	&19.2$^a$	&1554$^{+146}_{-123}$	&0.27&4.2$^{+0.8}_{-0.6}$	&7.8$^{+1.5}_{-1.2}$	&6.1$^{+0.6}_{-0.5}$ \\
RX\,J2133.7+5107		&0.95$\pm$0.04 & 7.14  	&570.82	&17.2$^a$	&1350$^{+49}_{-46}$	& 1.35 &18.7$^{+1.4}_{-1.2}$	&29.4$^{+2.2}_{-1.2}$	&9.3$\pm$0.3 \\
V418 Gem	&0.74$\pm$0.23 & 4.3704	&480.67	&11.2$^a$	&3842$^{+4444}_{-1341}$&0.24&47.7$^{+74.2}_{-27.5}$	&42.3$^{+155}_{-24}$	&4.8$^{+5.6}_{-1.7}$ \\
V515 And		&0.67$\pm$0.07 & 2.731	&465.5	&9.7$^a$			&1006$^{+49}_{-44}$	&0.77&13.2$^{+1.3}_{-1.1}$	&9.3$^{+	0.9}_{-0.8}$	&1.8$\pm$0.08 \\
V647 Aur		&0.85$\pm$0.16 & 3.46656	&932.9	&20.4$^a$		&2251$^{+402}_{-296}$	&0.37&18.4$^{+7.2}_{-4.5}$	&22.5$^{+8.7}_{-5.5}$	&10.2$^{+1.8}_{-1.3}$ \\
EI UMa		&0.91$\pm$0.08 & 6.4344	&741.6	&19.3$^a$	&1133$^{+50}_{-46}$	&0.72&7.8$^{+0.7}_{-0.6}$	&11.1$^{1.0}_{-0.9}$	&6.8$\pm$0.3 \\
V2069 Cyg	&0.83$\pm$0.10 & 7.48032	&743	&17.0$^a$	&1178$^{+46}_{-43}$	&0.53&7.7$\pm$0.6			&8.9$^{+0.7}_{-0.6}$	&4.7$\pm$0.2 \\
IGR\,J1719-4100	&0.72$\pm$0.06 & 4.0056	&1054	&18.3$^a$	&654.3$^{+17.9}_{-16.9}$	&1.39&8.1$^{+	0.5}_{-0.4}$ 	&7.1$\pm$0.4			&4.6$\pm$0.1 \\
IGR\,J0457+4527	&0.87$\pm$0.14 & 7.2	        &1223	&25.2$^a$	&4770$^{+14891}_{-2056}$	&0.48&99$^{+ 1589}_{-67}$	&130$^{+2072}_{-88}$	&36$^{+112}_{-16}$ \\
IGR\,J1817-2508	&0.54$\pm$0.08 & 1.5312	&1663.4	&18.5$^a$	&2130$^{+376}_{-278}$	&0.92&98$^{+38}_{-24}$		&50$^{+19}_{-12}$ 		&11.9$^{+2.1}_{-1.5}$ \\
IGR\,J0838-4831	&1.27$\pm$0.15 & 7.92	       &1480.8	&65.2$^a$	&2167$^{+327}_{-251}$	&0.24&3.2	$^{+1.0}_	{-0.7}$	&13.3$^{+4.3}_{-2.9}$	&89$^{+14}_{-10}$ \\
IGR\,J1509-6649	&0.85$\pm$0.09 & 5.8896	&809.42	&18.6$^a$	&1164$^{+39}_{-37}$	&0.67&8.9	$\pm$0.6			&10.8$\pm$0.7		&6.1$\pm$0.2 \\
IGR\,J1649-3307	&0.82$\pm$0.09 & 3.6168	&571.9	&14.2$^a$	&1170$^{+91}_{-79}$	&0.65&9.4	$^{+1.5}_{-1.2}$	&10.6$^{+1.7}_{-1.4}$	&3.8$\pm$0.3	 \\
IGR\,J1654-1916	&0.83$\pm$0.10 & 3.7152	&546.7	&14.2$^a$	&1096$^{+62}_{-56}$	&0.72&9.2$^{+1.0}_	{-0.9}$	&10.4$^{+1.2}_{-1.0}$	&3.7$\pm$0.2 \\
\hline
\end{tabular}
\begin{flushleft}{ 
Notes:  (a) Relative corotation radius; 
(b) Best fit model flux in the range 0.1--100 keV in units $10^{-10}$\,erg\,s$^{-1}$\,cm$^{-2}$;
(c) Distance assumed; there is not parallax measurement of XY Ari in Gaia DR2.
(*) X-ray spectra were fitted with the tall PSR models.
}\end{flushleft} 
\end{table*}

For some objects independent mass estimates are available, so it is
interesting to compare them with our results
(Fig.\,\ref{fig:opt}). In particular, we used mass estimates reported
by \citet{Hellier:97} for XY\,Ari, by \citet{Hellier:93} for
TV\,Col,\ by \citet{Haswell.etal:97} for DO\,Dra, by
\citet{Penning:85} for BG\,Cmi, by \citet{BR:08} for EX\,Hya, and  by
\citet{MR.etal:02} for GK\,Per. We note that in most cases the
agreement is excellent. The largest discrepancy is observed for
EX\,Hya where we still underestimate the mass, likely due to the fact
that our tall PSR model is actually still comparatively short
considering the observed accretion rate.

One could observe that Ritter's catalog of CVs \citep{RK:03} includes
more IPs with known WD masses. However, these estimates were also
obtained using X-ray observations and methods similar to one presented
here. It is not surprising, therefore, that for these objects the
agreement with our estimate is perfect if the same assumptions on PSR
geometry and magnetosphere size are used. We, thus, excluded
these objects from Fig.~\ref{fig:opt} for clarity. 

We can assess the properties of the WD mass distribution for
the entire IP population presented in Fig.\,\ref{fig:dstr}. It is
consistent with being normal (p-value of $\sim$\,0.2 for Shapiro-Wilk
test; \citealt{shapiro65}), and peaks around 0.7--0.9\,$M_\odot$,
which is consistent with past investigations
\citep{Zorotovic.etal:11}. The mean and standard deviation of WD mass in our distribution 
are $0.79\pm0.16 M_\odot$, which is close to existing estimates for CVs in
general by \citet{Zorotovic.etal:11} ($0.83\pm0.24 M_\odot$), and
specifically for IPs by \citet{Bernardini.etal:12} ($0.86\pm0.07
M_\odot$), and by \citet{Yuasa.etal:10} ($0.88\pm0.25 M_\odot$). All
these values are significantly larger compared to the isolated WD
population with 0.6\,$M_\odot$ \citep{Kepler.etal:16}. There are
hypotheses to explain this fact \citep[see discussion
  in][]{WZS:15}, but their consideration is out of scope of the
present work.

Gaia DR2 \citep{Gaia_mission,Gaiadr2} opened new possibilities for IP
studies. Now we have comparatively accurate determinations of the
distances to many IPs (see Table\,\ref{tab2}). Therefore, their
mass-accretion rates and surface magnetic fields can be estimated as
described above for EX\,Hya and GK\,Per. The results are presented in
Table\,\ref{tab2}. For IPs closer than 1\,kpc (i.e., excluding sources
with large distance uncertainties for clarity), the results are also
presented in Figs.\,\ref{fig:mdot} and \ref{fig:magn}.

 \begin{table}
\centering   
\caption{Comparison of WD magnetic field strengths in some IPs measured from
optical/IR polarization observations ($B_{\rm pol}^a$) with estimates obtained in the present work ($B_{\rm sp}$). 
 \label{tab:B} 
 }
\begin{tabular}[c]{l c c c c c }
\hline
Name & 	$M/$M$_\odot$        &  $B_{\rm pol}^a$,\,MG  &  $B_{\rm sp}$,\,MG   & $R_{\rm m}/R$        \\ 
\hline 
NY Lup		        &   1.05 	 	&   $>$ 4 		 &  	4.9  & 10.3 \\
BG CMi 			&    0.63 		&  $\sim$ 4	 &    4 	& 14.3$^b$     \\
V2731 Oph 		&  1.06 		&  $\sim$ 5	 &  	1.7 	& 7.8$^b$  \\ 
V2400  Oph		& 0.72 		&  9-20	         &  4.6	& 16.8$^b$  \\
IGR\,J1509 -6649	& 0.85		&   $>$ 10	 &  6.1 	& 18.6$^b$  \\
V2306 Cyg 		& 0.70		&  8			 &   7.8	& 22.1$^b$  \\
V405 Aur 			&  0.73 		& 32 		 &   2.2	& 12.0$^b$  \\
RX\,J2133+5107 	& 0.95	 	& $>$ 20		 &  9.3	& 17.2$^b$   \\
PQ Gem 			&  0.79 		&  8-21		 &  4.1	& 16.7$^b$      \\
 \hline
\end{tabular}
\begin{flushleft}{ 
Notes:  (a) References to the original works with the polarization measurements can be found in \citet{FMG:15}.
 (b) Relative corotation radius 
}\end{flushleft} 
\end{table}
 
It is interesting to look at the estimated mass-accretion rates
as a function of orbital period to facilitate comparison with existing
theoretical predictions \citep[see, e.g.][]{Howell.etal:01, GN:15}.
In particular, a comparison of our result with population synthesis
predictions (Fig.\,\ref{fig:mdot}) allows to conclude that most
IPs appear to have lower mass-accretion rates in comparison with the
predicted values. It seems also that the observed $\dot M$ decreases
with increasing orbital period. This behavior is opposite to
the expected one, but deviation is not significant. 
The mass accretion rate of GK\;Per during quiescence is on this 
dependence. We can suggest that most of IPs are in a low mass accretion rate
(quiescence) and might have  short, rare, and barely visible outbursts.  
We note,
however, that the mass-accretion rates of the IPs with orbital periods
less than four hours are close to the modeled values. Two IPs,
EX\,Hya and DO\,Dra, are in low states with observed mass-accretion
rates probably less than the average values. The WD masses in these
IPs were found using the tall PSR models. One short-period IP below
the period gap, V1025\,Cen, has a low mass accretion rate, close to the
expected value. The WD mass in this IP was also determined using the
tall PSR model spectra.

We compare the derived magnetic field strengths on the WD surface to
the orbital periods in Fig.\ref{fig:magn}. Our results range between 1
and 10\,MG, although in many cases the values shall be considered as
upper limits only, because we had to assume corotation for many IPs
due to lack of other magnetospheric radius estimates. The
actual magnetospheric radii might be smaller. The field strength in
EX\,Hya is extremely low, and this object is probably observed as an
IP only because of the extremely low mass-accretion rate, which raises
the question whether we miss a significant population of similar
sources which are too distant and faint for periodic variability to be
detected.

There are also several IPs with WD magnetic fields stronger than 10\,MG,
but all of them have distances beyond 2\,kpc, which are
poorly constrained. As a consequence, large magnetic field values might
just be due to overestimated distances.

The magnetic field strength can be also determined using
optical/IR polarization measurements \citep[see details in
][]{FMG:15}. These  authors provide a list of WDs in IPs investigated
in this way. We list them in Table\,\ref{tab:B} together with our
estimates. For three sources, namely NY\,Lup, V2306\,Cyg, and BG\,CMi
there is good agreement. On the other hand, for the rest of the
sources we significantly underestimate the field strength. We suggest that
this discrepancy arises due to the value of $\Lambda=0.5$ we assumed
above. In our case, the estimated field strength is
proportional to $\Lambda^{-7/4}$, so we might overestimate $\Lambda$
by a factor of 2--5 in some IPs. Note that the mass of the WD is
likely to be underestimated in these cases. On the other hand, the
value of $\Lambda$ has to be more or less the same in all IPs.
For NY\,Lup, where the magnetospheric radius has been estimated using
the break frequency in the power spectrum, the assumption of
$\Lambda=0.5$ gave an acceptable result.

Finally, it is important to emphasize that the magnetic field
estimates of some IPs (e.g., V405\,Aur) obtained using  polarization
measurements are extremely high, and comparable with values measured for polars \citep[see,
  e.g.,][]{FMG:15}. Existing PSR models are not applicable in
this case as cyclotron cooling dominates over thermal emission of the
optically thin non-magnetized plasma, making PSRs less bright in hard
X-rays compared to model prediction. However, we can still make
some conclusions based on available model. Indeed, these IPs with possibly high $B$ on the WD surface
are still bright in X-rays, therefore, we likely underestimate the WD mass in
these sources as well as the mass-accretion rates and the magnetic
field strengths if the polarization-based estimates of the magnetic field are correct.

\section{Summary}
\label{sec:summary}

We conducted a systematic analysis of a large sample of
35 IPs with the aim to estimate WD masses in these sources. In
particular, we used a new PSR model taking into the account the finite
magnetospheric radius to describe their hard X-ray spectra. For many
sources we were able to obtain an independent estimate of the
magnetosphere size based on the observed frequency break in the power
spectrum of their aperiodic variability.

Two 2-parametric spectral model grids, with WD mass $M$ and relative
magnetospheric radius $R_{\rm m}/R$ as free parameters, were computed
under two limiting assumptions. The first grid assumes a high local
mass-accretion rate and short PSRs, and it is similar to earlier
works, and has been published before \citep{Suleimanov.etal:16}. A
second grid with a fixed, tall relative PSR height of $H_{\rm
  sh}/R=0.25$ was computed in the present work. 

We used archival spectra of IPs measured with the NuSTAR and Swift/BAT
observatories. We only considered the hard part of the spectra ($E>
20$\,keV) to avoid problems associated with the treatment of complex
absorption and/or reflection in the soft band. For sources observed
by NuSTAR we also carried out timing analysis to constrain the break
frequency. Here we assumed that the break frequency corresponds to the
Keplerian frequency at the inner edge of the accretion disc disrupted by
the rotating magnetosphere. In cases where no break could be detected,
either due to insufficient data quality or absence of timing data, we
assumed that the magnetospheric radius is comparable with the
corotation radius of a given source.

We considered the largest sample of IPs to date, and obtained the
largest sample of uniformly deduced WD mass estimates. The resulting
mass distribution is consistent with being normal with an average of
0.79\,$M_\odot$ and mean dispersion of 0.16\,$M_\odot$, i.e., slightly
lower values compared to previous works. 

Using distances recently made available by Gaia DR2, we obtained
for the first time robust estimates of mass-accretion rates for the
investigated IPs. Most sources accrete at comparable rate
around 10$^{-9}$\,M$_\odot$\,yr$^{-1}$. Two IPs below the period gap,
EX\,Hya and V1025\,Cen, accrete at significantly lower rate, $<
10^{-10}$\,M$_\odot$\,yr$^{-1}$. In the case of EX\,Hya, as well as
DO\,Dra, this is likely due to the fact that these sources are dwarf
novae currently being in low state with depressed mass-accretion rate. 

Using the obtained results for mass-accretion rates and magnetospheric
radii, we evaluated WD surface magnetic field strengths for several
objects. We found them to be in range 1--10\,MG in most cases. The
unusual object EX\,Hya appears to have a significantly weaker field of
about 10$^4$\,G. We compared our field strength estimates with values
published in literature based on optical/IR polarization measurements
\citep[nine IPs, see][]{FMG:15}, and find good agreement for three
sources, whereas for the other six our results are smaller by
factors of 2--15. This discrepancy might indicate that  we overestimate the ratio of
the magnetospheric radius to the Alfv\'{e}n radius in these IPs ($\Lambda <0.5$), and 
we have to take into account the cyclotron cooling in our models to describe PSRs
at highly magnetized WDs in these IPs.

Two IPs with extremely small magnetospheres, EX\,Hya and GK\,Per, were
investigated in more detail. We showed that a tall PSR is
necessary to explain the observed hard X-ray spectrum of the
low-luminosity IP EX\,Hya. We also investigated the magnetosphere size of
the dwarf nova GK\,Per in outburst and in quiescence, and
determined mass-accretion rates in both states. For the first time we
were able to detect a break in the power spectrum of this source
during quiescence, and to determine the break frequency and magnetosphere
size in both states. We found that the magnetosphere size changes
between the different luminosity states as expected from the 
Alfv\'{e}n law $R_{\rm m} \sim \dot M^{-2/7}$ within 2$\sigma$.

\section*{Acknowledgments} 
This work has made
use of data from the European Space Agency (ESA) mission {\it Gaia}
(\url{https://www.cosmos.esa.int/gaia}), processed by the {\it Gaia}
Data Processing and Analysis Consortium (DPAC,
\url{https://www.cosmos.esa.int/web/gaia/dpac/consortium}). Funding
for the DPAC has been provided by national institutions, in particular
the institutions participating in the {\it Gaia} Multilateral
Agreement.
The work was supported by the German Research Foundation (DFG) grant
WE 1312/51-1,  and  the Russian Foundation for Basic Research  grant
{18-42-160003 r\_a} (VFS). VD thanks the Deutsches Zentrum for Luft-
und Raumfahrt (DLR) and DFG for financial support. 



\bibliographystyle{mnras}
\bibliography{ip} 

\bsp	
\label{lastpage}
\end{document}